\documentclass{aa}
\usepackage{graphicx}
\usepackage[latin1]{inputenc}
\usepackage{lscape}
\usepackage{aalongtable}
\usepackage[figuresright]{rotating}

\newcommand{\kms}   {~km~s$^{-1}$}
\newcommand{\cmd}   {~cm$^{-2}$}
\newcommand{\cmt}   {~cm$^{-3}$}

\newcommand{\nh}    {NH$_3$}

\newcommand{\hco}   {HCO$^+$}
\newcommand{\dco}   {DCO$^+$}
\newcommand{\htco}  {H$^{13}$CO$^+$}

\newcommand{\htcs}  {H$_2$CS}
\newcommand{\for}   {H$_2$CO}
\newcommand{\met}   {CH$_3$OH}
\newcommand{\cthd}  {C$_3$H$_2$}
\newcommand{\cthc}  {C$_3$H$_4$}
\newcommand{\cdo}   {C$^{18}$O}
\newcommand{\tco}   {$^{13}$CO}
\newcommand{\J}[2]  {\mbox{#1--#2}}
\newcommand{\JK}[4] {\mbox{#1$_{#2}$--#3$_{#4}$}}

\newcommand{\ap}    {A$^{+}$}

\newcommand{\N}[1]  {$\times10^{#1}$}

\newcommand{\cmcd}   {~cm$^{-2}$}
\newcommand{\cmgd}   {~cm$^{-3}$}

\begin{document}
\title{An observational survey of molecular emission ahead of Herbig-Haro
objects}
\author{Serena Viti\inst{1}, 
Josep Miquel Girart\inst{2,3}, Jennifer Hatchell\inst{4}}
\offprints{S. Viti: sv@star.ucl.ac.uk}
\institute{Department of Physics and Astronomy, UCL, Gower St., London,
WC1E~6BT, UK  
\and 
Institut de Ci\`encies de l'Espai (CSIC), Gran Capit\`a 2, 08034 Barcelona,
Catalunya, Spain \and Institut d'Estudis Espacial de Catalunya, Spain  
\and
School of Physics, University of Exeter, Stocker Road, Exeter, EX4 4QL, UK 
}
\date{Accepted ... Received ...}
\abstract
{A molecular survey recently performed ahead of HH~2
supports the idea that the observed molecular enhancement is due to UV
radiation from the HH object.}
{The aim of the present work is to
determine whether all HH objects with enhanced HCO$^+$ emission ahead
of them also exhibit the same enhanced chemistry as HH~2.
We thus observed several molecular lines at several positions ahead
of five Herbig-Haro objects where enhanced HCO$^+$ emission was
previously observed.}
{We mapped the five Herbig-Haro objects using the IRAM-30 m. For
each position we searched for more than one molecular species, and
where possible for more than one transition per species. We then
estimated the averaged beam column densities for all species observed
and also performed LVG analyses to constrain the physical properties
of the gas.}
{The chemically richest quiescent gas is found ahead of the HH~7-11
complex, in particular at the HH~7-11 A position. In some regions we
also detected a high velocity gas component.  We find that the gas
densities are always higher than those typical of a molecular cloud
while the derived temperatures are always quite low, ranging from 10
to 25 K. The emission of most species seems to be enhanced with
respect to that of a typical dense clump, probably due to the exposure
to a high UV radiation from the HH objects. Chemical differentiation
among the positions is also observed. We attempt a very simple
chemical analysis to explain such differentiation.} {}

\keywords{
ISM:  Herbig-Haro objects ---
ISM: abundances ---
ISM: clouds ---
ISM: molecules ---
Radio lines: ISM ---
Stars: formation
}

\authorrunning{S. Viti et al.}
\titlerunning{A molecular survey ahead of HH objects}
\maketitle
\section{Introduction}

Herbig-Haro (HH) objects are regions where protostellar jets interact
hydro-dynamically with the surrounding molecular cloud, resulting in
strong atomic line emission characteristic of shocks.  There is now
ample and increasing evidence that HH objects also have a less direct
but very diverse chemical influence on nearby molecular clumps which
have not yet been reached completely by the flow itself. In fact, the
radiation generated in the HH objects, as well as the shocks produced
by the HH objects alter the chemical composition of the molecular gas
surrounding them in a very complex way.  An example of the complex
interaction of the HH object with its surroundings, and
the most studied one, is the 
HH~2 region (e.g. \cite{girart02}; \cite{dent03};
\cite{lef05}; \cite{girart05}). Some of the gas ahead of the HH object
seems to be altered solely by the UV radiation: in the last decade or
so quiescent, cold, molecular condensations have been found ahead of
several HH objects. These cool condensations were found in enhanced
emission of HCO$^+$ and NH$_3$, which are tracers of high density gas
(e.g. \cite{rudolph88}; \cite{torrelles92}; \cite{girart94}). These
emissions probably arose because the UV radiation generated in the HH
shock evaporated the icy mantles on dust grains in small density
enhancements in the cloud near to the head of the jet
(\cite{girart94}).  Theoretical and further observational studies
(\cite{vw99}, hereafter VW99; \cite{viti03} and \cite{girart05})
confirmed this picture and showed that, because of the quiescent
nature of such clumps (clear from the linewidths of the emission as
well as the derived kinetic temperature of the gas), the high
abundances of HCO$^+$ and other species could not be due to shock
chemistry.

In this paper, we are particularly interested in these anomalous,
quiescent enhanced molecular emissions from localized regions in the
vicinity of the HH objects.  The VW99 models succeeded in explaining
the known enhancements of HCO$^+$ and NH$_3$ relative to CS, predicted
the amount by which these species should be increased, but also
indicated other species which should be enhanced (by factors of
10--200 with respect to dark molecular clouds e.g L1554), in
particular CH$_3$OH, H$_2$S, SO, SO$_2$, H$_2$CO, C$_3$H$_4$, H$_2$CS.
Such predictions were confirmed by recent observations and further
modelling of HH~2 (\cite{girart02}, \cite{viti03};
\cite{girart05}). It is important to note however that while single
dish observations (\cite{girart02}, \cite{viti03}) show that the
chemistry of the illuminated $quiescent$ clumps ahead of the HH~2
object is essentially consistent with the UV field from the HH~2 front
releasing the grain mantles and hence enhancing the gas chemistry,
BIMA maps of such region in several molecules expected to be
enhanced (\cite{girart05}) have shown how the HH~2
object affects the chemistry and dynamic of the surrounding
environment in different degrees. In summary, although Girart et al. (2005)
concluded that a very complex morphological kinematical and chemical
structure of the molecular gas is present ahead of the HH~2 object, 
%
they also confirmed that there is indeed an apparently quiescent
molecular component that is more exposed to the UV radiation from the
HH~2 front; the analysis of observational molecular lines from this
region can be explained by UV radiation from the HH shock lifting the
ice mantles from grains of a multi-density component clump and the
consequent chemistry that arises is as complex and varied as indicated
in the VW99 models.

The question we attempt to answer in this paper is the following: is
the rich photochemistry characteristic of UV illumination peculiar to
the HH~2 or is it common to all HH objects?  The simplest and most
direct way to answer such question is to perform a survey of molecular
species predicted to be enhanced by the VW99, ahead of several HH
objects where enhanced quiescent HCO$^+$ and/or NH$_3$ have already
been detected. We have now performed such a survey with the IRAM
30-meter and we report the observational results in this paper. The
main aims of this paper are: 1) determine whether other HH objects
exhibit the same peculiar enhanced chemistry of HH~2 as predicted by
VW99 and 2) assuming the HH object to be the source of the UV
radiation, look for correlations between the enhanced molecular
emission and the distance of such emission to the HH object.

We report our observations in \S~2; in \S~3 we present our results and
the derivation of beam averaged column densities. In \S~4 we analyse
our data by means of radiative transfer modelling. In \S~5 we briefly
discuss the chemical characteristics of the gas and in \S~6 we
summarize our findings.

\section{Observations}

The observations were carried out with the IRAM 30 m radio telescope
on Pico Veleta in November 2000. Our sample consisted of five Herbig
Haro regions; for most of them we took several positions where
enhanced HCO$^+$ was previously detected (see individual subsections
and Table~\ref{tb:pos}).  For each position we searched for several
molecular emissions; when possible we searched for more than one
transition of the same species (see Table~\ref{tb:lines}).  The
molecular species chosen are those which were expected from VW99
models to be enhanced (with respect to dark molecular clouds) by
reasonably large factors, along with tracers of the underlying column
density in the cloud (CS and CO isotopes) and tracers of the
ionization (the ratio DCO$^+$/HCO$^+$; \cite{caselli98}). Due to time
constraints, we could not attempt to detect all the species predicted
by the VW99 at every position. Our criteria varied from object to
object but in general if HCO$^+$ emission was weak we did not attempt
a detection in other species. We also prioritized species easier to
observe, so for example, although both SO and SO$_2$ are predicted by
the models to be enhanced we concentrated on SO as its lines are
usually stronger and also it has a higher predicted abundance. All the
transitions were observed in the frequency bands of 100, 150, 230 and
270 GHz (see Table 2 for line frequencies).  Integration times
were typically 2-12 minutes (position switched), with longer attempts
at detecting weak lines (e.g. NS: 30~mins).  The spectral resolution
was 40 kHz (100/150 GHz bands) or 80 kHz (230/270 GHz bands) resulting
in a velocity resolution typically of 0.05~km/s and at most 0.13~km/s.
The bandwidth was at least 20 MHz (100/150 GHz bands) or 40 MHz
(230/270 GHz bands).  Conditions were generally poor and variable for
winter nights with 1--6mm PWV (precipitable water vapour).  Resulting
system temperatures were T${\rm sys} \sim 100\hbox{--}120$, 200--460,
200--300 and 400-700~K for the 100, 150, 230 and 270~GHz bands
respectively, depending on frequency and weather.  Pointing was good
to within 4$\\arcsec$.  Spectra were corrected for the beam
efficiencies $B_{\rm eff} = 0.75$, 0.69, 0.51, and 0.42 in the 100,
150, 230, and 270~GHz bands respectively. The 30m beam FWHM ranges as
$\nu^{-1}$ from $28''$ at 86~GHz down to $11''$ at 230~GHz. See
on-line Tables~\ref{tb:obshh7a} to ~\ref{tb:obsngc} for details of the
molecular transitions observed.

\begin{table}
\caption{Source list}
\begin{tabular}{|c|ccc|}
\hline
Object & RA & DEC & V$_{LSR}$ \\
 & J2000 & J2000 & kms$^{-1}$ \\
\hline
HH~7--11 A   &03:29:06.5 &   31:15:35.5 & +8.0\\
HH~7--11 B   &03:29:06.9 &   31:15:47.4 & +8.0\\
HH~7--11 X   &03:29:05.7 &   31:15:30.5 & +8.0\\
HH 1         &05:36:20.2 &$-$06:45:01.3 & +9.0\\
HH 34 A      &05:35:34.3 &$-$06:29:07.0 & +8.5\\
HH 34 C      &05:35:32.2 &$-$06:29:04.9 & +8.5\\
HH 34 D      &05:35:32.0 &$-$06:29:23.9 & +8.5\\
HH 34 E      &05:35:30.6 &$-$06:29:03.8 & +8.5\\
JMG 99 G1    &06:41:04.5 &   09:56:02.0 & +4.6\\
JMG 99 G2    &06:41:07.1 &   09:56:07.0 & +4.0\\
\hline
\end{tabular}
\label{tb:pos}
\end{table}

The choice of our sample of objects was made by selecting all the HH objects
with known HCO$^+$ emission enhancements.  For some objects, we detected two
velocity components in several species: a high velocity one, probably coming
from the closest part to the HH object where the gas has already been
dynamically affected; and an ambient one. In this paper we will mainly 
concentrate on
the latter.

     \begin{table*}
     \caption[]{Frequency setups of the IRAM 30m observations and list of
positions observed.}
     \label{tb:lines}
     \[
     \begin{tabular}{lrrcccc@{\hspace{0.3cm}}ccccc@{\hspace{0.3cm}}cccc}
     \noalign{\smallskip}
     \hline
     \noalign{\smallskip}   
\multicolumn{1}{c}{Molecular} &
\multicolumn{1}{c}{Frequency} &
\multicolumn{1}{c}{Beam} & 
\multicolumn{4}{c}{HH~7--11} &
\multicolumn{5}{c}{HH 34} &
\multicolumn{2}{c}{$\!\!\!\!\!$N2264G$^a$} &
\multicolumn{1}{c}{}
\\
\cline{4-6}\cline{8-11}\cline{13-14}
\multicolumn{1}{c}{Transitions} & 
\multicolumn{1}{l}{(GHz)} &
\multicolumn{1}{l}{($''$)} 
                                        & A& X& B&& A& E& C& D&& 1& 2& HH 1 \\
     \noalign{\smallskip}     \hline
     \noalign{\smallskip}      
\tco\ \J{2}{1}            &220.3980& 11.0& X& X& X&& X & X& X& X&& X& X& X \\
\cdo\ \J{2}{1}            &219.5604& 11.0& X& X&  && X & X&  &  && X& X& X \\
\hco\ \J{1}{0}            & 89.1885& 28.0& X& X& X&& X & X& X& X&& X& X& X \\
\hco\ \J{3}{2}            &267.5576&  9.0& X& X& X&& X & X& X& X&& X& X& X \\
\htco\ \J{1}{0}           & 86.7543& 29.0& X& X& X&& X & X& X& X&& X& X& X \\
\htco\ \J{3}{2}           &260.2555&  9.5& X& X& X&& X & X&  &  && X& X& X \\
\dco\ \J{2}{1}            &144.0773& 17.0& X& X&  && X & X&  &  && X& X& X \\
\dco\ \J{3}{2}            &216.1126& 12.0& X& X&  && X & X&  &  && X& X& X \\
HCN \J{3}{2}              &265.8864&  9.0& X& X&  && X & X&  &  && X& X& X \\
\for\ \JK{2}{0,2}{1}{0,1} &145.6029& 17.0& X& X&  && X & X&  &  &&  &  & X \\
\for\ \JK{3}{1,3}{2}{1,2} &211.2115& 12.0& X&  &  &&   & X&  &  &&  &  &   \\
\met\ \JK{2}{k}{1}{k}$^b$ & 96.7414& 26.0& X& X&  && X & X&  &  && X& X&   \\
\met\ \JK{3}{k}{2}{k}$^c$ &145.1032& 17.0& X& X&  && X & X&  &  && X& X&   \\
\met\ \JK{5}{k}{4}{k}$^d$ &241.7914& 10.0& X& X& X&&   &  &  &  &&  &  &   \\
\cthc\ \JK{5}{0}{4}{0}    & 85.4573& 29.0& X&  &  &&   & X&  &  &&  &  &   \\
\cthd\ \JK{2}{1,2}{1}{0,1}& 85.3389& 29.0& X&  &  &&   & X&  &  &&  &  &   \\
\cthd\ \JK{3}{1,2}{2}{2,1}&145.0896& 29.0& X& X&  &&   &  &  &  &&  &  &   \\
\cthd\ \JK{4}{1,4}{3}{0,3}&150.8519& 17.0& X&  &  &&   &  &  &  &&  &  &   \\
CS \J{3}{2}               &146.9690& 17.0& X& X& X&& X & X& X& X&& X& X& X \\
CS \J{5}{4}               &244.9356& 10.0& X&  &  &&   & X&  &  &&  &  &   \\
SO \JK{3}{2}{2}{1}        & 99.2999& 20.0& X& X&  && X & X&  &  && X& X& X \\
SO \JK{6}{5}{5}{4}        &219.9494& 11.5& X& X&  && X & X&  &  && X& X& X \\
SO \JK{7}{6}{6}{5}        &261.8437&  9.5& X& X&  && X & X&  &  && X& X& X \\
H$_2$S \JK{1}{1,0}{0}{0,1}&168.7628& 14.5& X& X& X&&   &  &  &  && X& X& X \\
\htcs\ \JK{4}{0,4}{3}{0,3}&137.3711& 18.0& X&  &  &&   & X&  &  &&  &  &   \\
HCS$^+$ \J{5}{4}          &213.3606& 12.0& X&  &  &&   &  &  &  &&  &  &   \\
NS \J{7/2}{5/2}           &161.2972& 15.0& X&  &  &&   & X&  &  &&  &  &   \\
OCS \J{7}{6}              & 85.1391& 29.0& X&  &  &&   &  &  &  &&  &  &   \\
     \noalign{\smallskip}
     \hline
     \end{tabular}
     \]
     \begin{list}{}{}
\item[$^a$] Abbreviation of NGC 2264G JMG1 and JMG2
\item[$^b$] \JK{2}{0}{1}{0} \ap\ transition. 
\item[$^c$] \JK{3}{0}{2}{0} \ap\ transition. 
\item[$^d$] \JK{5}{0}{4}{0} \ap\ transition. 
     \end{list}
    \end{table*}
%

%

%
%

\begin{table*}
\caption{Averaged beam column densities of the low velocity component in the
clumps ahead of HH objects observed with IRAM 30m (Columns 2-10); the last column shows, for reference, the column densities (when available) at position I3 ahead of HH2 taken from Girart et al. (2005, table A.2). To fit the table, we abbreviated HH~7--11 to 7-11, the HH 34 clumps to HH34, and NGC~2264~G JM G2 to JMG~99~G2. a(b) stands for a$\times$10$^b$.}
\begin{tabular}{|c|c|c|c|c|c|c|c|c|c|c|}
\hline 
\textbf{MOL.}
 &\textbf{7-11A}
 &\textbf{7-11X}
 &\textbf{7-11B}
 &\textbf{HH34 A}
 &\textbf{HH34 E}
 &\textbf{HH34~C}
 &\textbf{HH34~D}
 &\textbf{HH~1}
 &\textbf{JMG~99~G2}
 &\textbf{HH 2}
 \\
 \hline 
\textbf{CO$^a$}    &2.6(18) &2.3(18) &2.1(18) &1.9(18) &1.4(18) &1.3(18) &1.2(18) &3.5(18) &3.7(17) & 3.3(17)\\
 \hline 
\textbf{$^{13}$CO} &4.6(16) &4.0(16) &3.7(16) &3.0(16) &2.2(16) &2.0(16) &1.9(16) &5.5(16) &6.5(15)& -- \\
 \hline 
\textbf{\met}      &3.5(14) &4.1(14) &---    &2.9(13) &4.9(13) &---    &---    &5.6(13) &9.8(13) & 6.2(13)\\
 \hline 
\textbf{\hco$^b$}  &2.0(14) &1.6(14) &2.2(14) &1.2(13) &3.3(13) &1.7(13) &9.7(12) &2.1(13) &5.0(12) & 4.0(13)\\
 \hline 
\textbf{\htco}     &2.3(12) &1.8(12) &2.5(12) &1.5(12) &4.2(12) &---    &---    &2.6(12) &6.0(11) & -- \\
 \hline 
\textbf{SO}        &8.5(13) &8.6(13) &---    &8.2(12) &1.7(13) &---    &---    &1.7(13) &1.6(13) & 1.7(13) \\
 \hline 
\textbf{\for$^c$}  &9.3(13) &9.6(13) &---    &1.0(13) &1.1(13) &---    &---    &1.6(13) &---  & 9.0(13)  \\
 \hline 
\textbf{CS}        &6.0(13) &6.2(13) &5.9(13) &6.8(12) &8.2(12) &6.2(12) &5.6(12) &1.0(13) &1.1(13) & 2.0(12)\\
 \hline 
\textbf{H$_2$S}    &5.0(13) &4.7(13) &4.5(13) &---    &---    &---    &---    &1.2(13) &6.0(12) & --  \\
 \hline 
\textbf{C$_3$H$_4$}&5.7(13) &---    &---    &---    &$<$2(13)&---    &---    &---    &---   & -- \\
 \hline 
\textbf{HCN}       &2.3(13) &1.6(13) &---    &$<$6(11)&$<$6(11)&---    &---    &1.8(12) &7.8(12) & 2.8(12) \\
 \hline 
\textbf{OCS}       &2.1(13) &---    &---    &---    &---    &---    &---    &---    &---  & --  \\
 \hline 
\textbf{H$_2$CS}   &1.1(13) &---    &---    &---    &$<$2(12)&---    &---    &---    &---  & --  \\
 \hline  
\textbf{C$_3$H$_2$}&8.9(12) &8.4(12) &---    &---    &8.5(11) &---    &---    &---    &---  & --  \\
 \hline 
\textbf{NS}        &7.8(12) &---    &---    &---    &$<$5(12)&---    &---    &---    &---   & --  \\
 \hline 
\textbf{DCO$^+$}   &5.0(12) &3.6(12) &---    &3.8(11) &2.9(11) &---    &---    &1.9(11) &2.2(11) & 7.8(11) \\
 \hline 
\textbf{HCS$^+$}   &$<$2(12)&---    &---    &---    &---    &---    &---    &---    &--- & --  \\
 \hline 

\end{tabular}
     \begin{list}{}{}
\item[$^a$] Optical depth of HH~7--11~B adopted from HH~7--11~X, and of  HH 34~C
 and HH 34~D adopted from HH 34 E. 
\item[$^b$] Optical depth of HH 34~C and HH 34~D adopted from HH 34 E. 
\item[$^c$] Adopted an ortho/para ratio of 1.5
     \end{list}
\label{tb:cd}
\end{table*}

\subsection{Brief description of the sources}

\subsubsection{HH~7--11}

%
%
     \begin{figure}
     \resizebox{8cm}{!}{\includegraphics{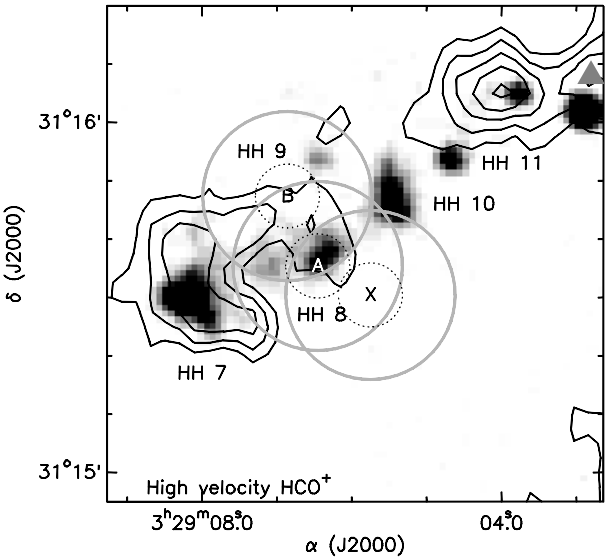}}
     \hfill
     \caption[]{Sketch of the HH~7--11 region. The gray scale shows the optical
R  band image of the HH~7--11 objects, while the contour plots shows the BIMA 
HCO$^+$ 1--0
blueshifted high velocity emission (from Mehringer 1996). The optical image was taken 
from the public catalog Aladin. The circles are the IRAM 1mm (dotted) and 
3mm (solid) beams at the A, B and X positions.
}
     \label{fhh7a}
     \end{figure}
%

The HH~7-11 system (d $\sim$220 pc: \cite{cernis90}) consists of
fairly low excitation HH objects, with HH~7 being at the far end of
the outflow (\cite{solf87}).  However, the region surrounding HH~7--11
is very rich in young stellar objects (e.g. Rodr{\'{\i}}guez et
al. 1999), hence the total radiation impinging on the surroundings is
unknown. Enhanced HCO$^+$ \J{1}{0}\ emission has been previously
observed just ahead of HH~8--10 objects in the form of small clumps
with 7$''$ sizes (1500~AU, \cite{rudolph88}).  CS and NH$_3$
(\cite{rudolph01}) are also reported towards these clumps.
Submillimeter observations in the region reveal a dust source, HH
7--11 MMS 4, that engulfs the HCO$^+$ clumps (\cite{chini01}). From
the dust observations Chini et al. (2001) derive that the total mass
of the quiescent gas around HH~8--10 is 4.0~M$_{\odot}$. From the
study of the HCO$^+$ \J{3}{2}\ emission, Dent et al. (1993) suggest
that the dense molecular gas and dust around HH~8--11 is the densest,
slowest component of the outflow. This is somewhat similar to the
picture of the molecular gas ahead of HH~2 (\cite{girart05}).

We selected three positions, two of them, HH~7--11 A and B, toward the \hco\
\J{1}{0}\ clumps (\cite{rudolph88}) and one, HH~7--11 X, toward a
HCO$^+$ \J{3}{2}\ emission peak southwest of HH~8 (\cite{dent93}). Note that
the 3 mm beams of our 3 positions (A, X and B) intersect (see
Figure~\ref{fhh7a}).

\subsubsection{HH 1}

\begin{figure}
     \resizebox{8cm}{!}{\includegraphics{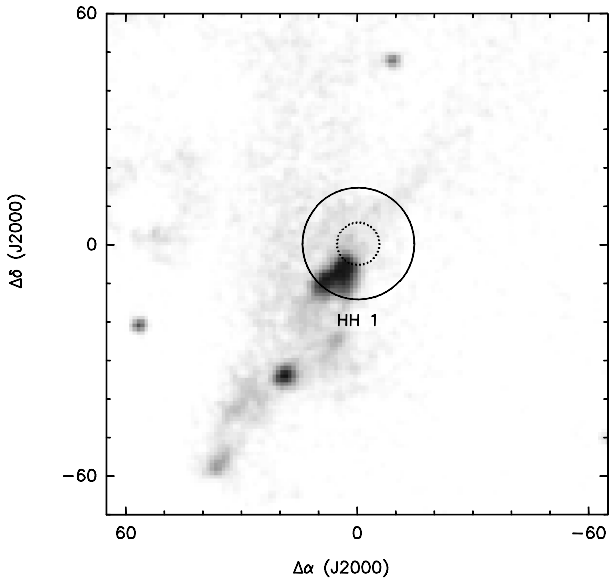}}
     \hfill
     \caption[]{          
Sketch of the HH~1 region. The gray scale shows the optical R band image
of the HH~1 object, taken from the public catalog
Aladin. The circles are the IRAM 1mm (dotted) and 3mm (solid) beams.
}
     \label{fhh1} 
     \end{figure}

The HH 1--2 system (d $\sim$450 pc) is one of the most intensely
studied HH object system to date. The proper motions of HH~1 and HH~2
show that they are moving in opposite directions (\cite{hj81}). Their
powering source is probably the embedded luminous young star known as
HH~1--2 VLA~1 (\cite{pravdo85}). Infrared and optical images show very
complex structures, quite randomly positioned in the case of HH~2 and
more like a set of bow shock features in HH~1. Clumpy molecular
emission ahead of HH~2 has been the subject of an extensive recent
study (\cite{girart05}).  Enhanced \hco\ and \nh\ have also being
previously observed ahead of HH~1 (\cite{davis90};
\cite{torrelles93}).  No dust is detected towards the \nh\ clumps
(\cite{chini01}) up to 0.22~Jy~beam$^{-1}$ (3-$\sigma$) at
850~$\mu$m. Note that the dust peak emission ahead of HH~2 is
0.20$\pm0.04$~Jy (\cite{dent03}). Our IRAM 30 m observations pointed
toward the \nh\ (1,1) ammonia peak ahead of HH~1 (see
Figure~\ref{fhh1}). We did a search for the most abundant species
towards this region.

%
%

\subsubsection{HH~34}

\begin{figure}
     \resizebox{8cm}{!}{\includegraphics{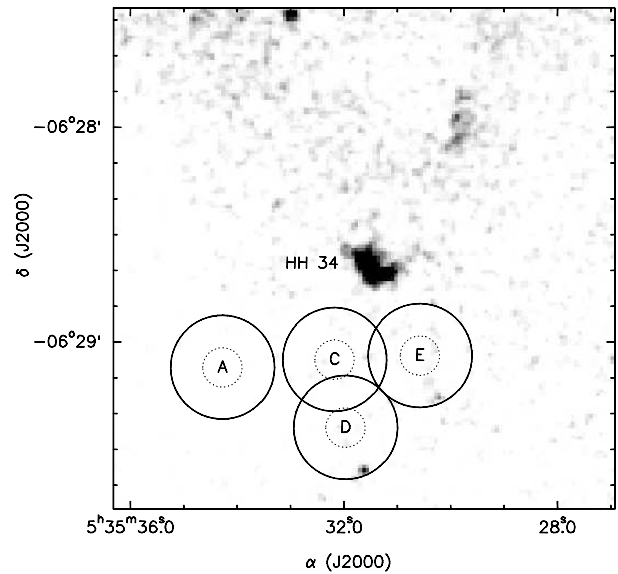}}
     \hfill
     \caption[]{          
Sketch of the HH~34 region. The grey scale shows the optical R band image of
the HH 34 object, taken from the public catalogue Aladin. The circles are the 
IRAM 1mm (dotted) and 3mm (solid) beams at the HH~34 A, C, D and E positions.
}
     \label{fhh34} 
     \end{figure}

The HH 34 is a high excitation large bow shock (\cite{reipurth02}),
which is part of a parsec scale HH system located in the molecular
cloud L1641 in Orion (d$\sim$480 pc). Its source may be a faint star
at 46,000 AU away.  Enhanced, and quiescent, HCO$^+$ \J{1}{0}\
emission was detected from five clumps with sizes of about 15$''$ or
7000~AU (\cite{rudolph92}). The clumps are located south of HH~34 at a
distance to the HH object of 0.04-0.09 pc, corresponding to
8000--18000 AU.  We have searched for the predicted set of molecules
at four positions which coincide with the peak \hco\ emission of four
of the detected clumps, A, E, C and D, by Rudolph et
al. (1992; see Figure~\ref{fhh34}), although for these two last
positions only \tco, \hco\ and CS were observed.

%
%

\subsubsection{NGC 2264G}

NGC~2264G is associated with a spectacular CO outflow driven by a
Class 0 protostar (\cite{lada96}; \cite{gomez94};
\cite{ward95}). Girart et al. (2000) found two small clumps (sizes of
$\la 14''$) of apparently quiescent HCO$^+$ \J{4}{3} emission, JMG 99
G1 and JMG 99 G2, west of the protostar and just ahead from
shock-excited near-IR emission.  Recent observations shows that these
clumps are also associated with strong high velocity SiO and
CH$_{3}$OH emission (\cite{garay02}). We observed these two clumps.

%
     \begin{figure}
     \resizebox{8cm}{!}{\includegraphics{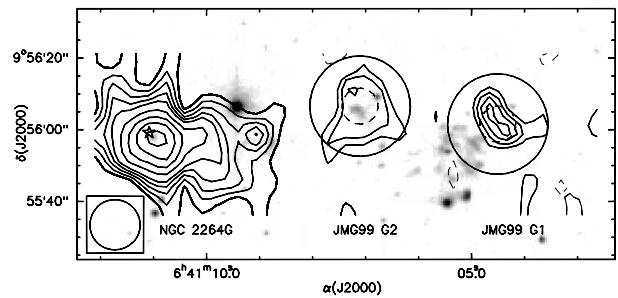}}
     \hfill
     \caption[]{          
JCMT contour maps of the integrated HCO$^+$ \J{4}{3} emission (from Girart  et
al. 2000) overlapped with the grey scale image of the near-IR H$_{2}$ emission.
The circles are the IRAM 1mm (dashed) and 3mm (solid) beams at the
JMG99 G2 and G1 positions.
}
     \label{fngc} 
     \end{figure}
%

\section{Results}

\begin{figure*}
     \rotatebox{-90}{\resizebox{10.2cm}{!}{\includegraphics{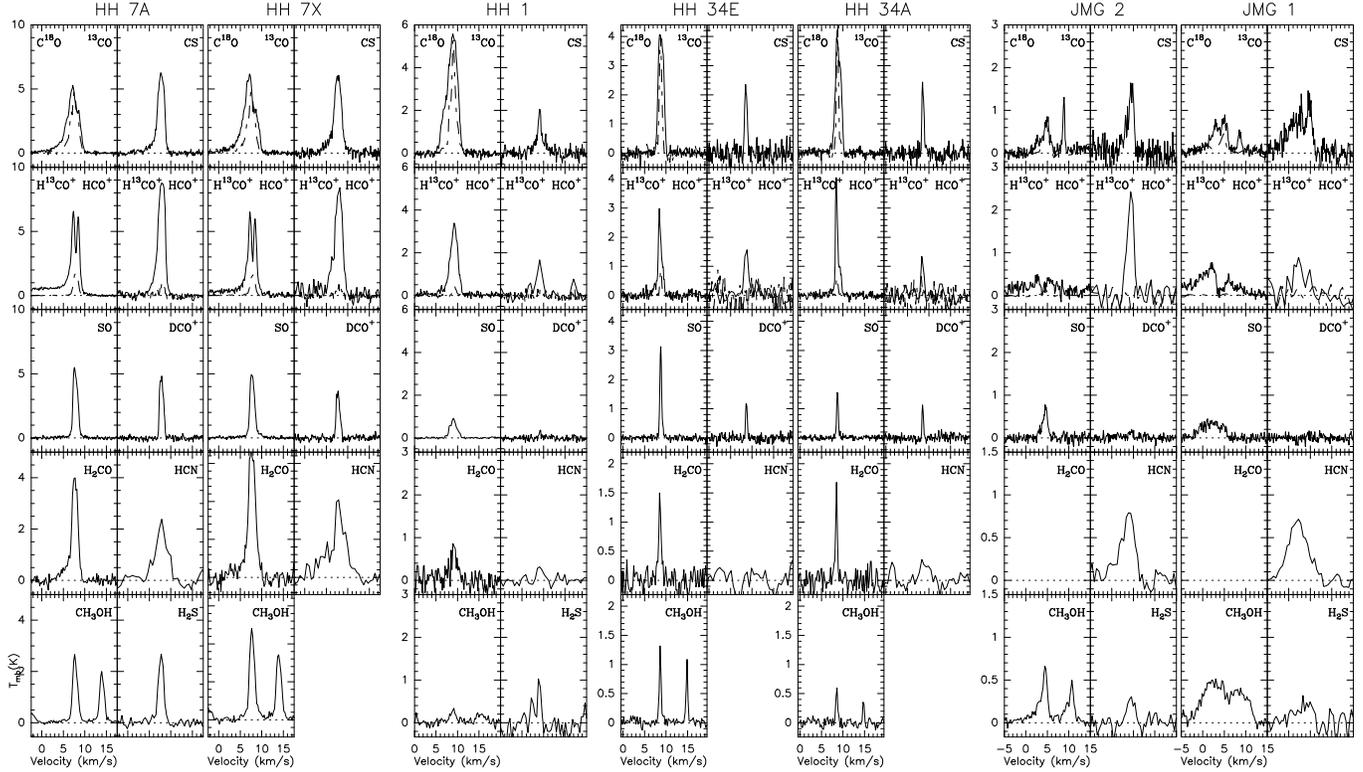}}}
     \hfill
     \caption[]{Spectra of selected transitions of the different species
observed  toward 
(from left to right) HH~7--11~A,  HH~7--11X, HH~1, 
HH~34E, HH~34A, JMG~1 and JMG~2.
\tco\ \J{2}{1}\ spectra is scaled down by a factor of 3 to fit in the boxes.
The transition lines showed are (from left to right and from top to bottom):
\tco\ 2--1, CS 3--2, \hco\ 1--0, \hco\ 3--2, SO \JK{3}{2}{2}{1}, 
DCO$^+$ 2--1, H$_2$CO \JK{2}{0,2}{1}{0,1}, HCN 3--2, 
\met\ \JK{2}{K}{1}{K}, and H$_2$S \JK{1}{1,0}{0}{0,1}.
The dashed lines in the CO and HCO$^+$ panels are for the
C$^{18}$O  and HC$^{13}$CO$^+$ isotopes at the same transition, respectively.
}
     \label{fhh7ahh34e} 
     \end{figure*}

Figure~\ref{fhh7ahh34e} shows examples of the spectra of the molecular
transitions for which emission was detected, for HH~7--11~A and HH~34 E.
HH~7--11 is the region where molecular lines are brighter so position HH~7--11~
A was selected to carry out the most complete survey. Apart from HCS$^+$, all 
the molecular species predicted by the VW99 to be enhanced were detected
toward this position. All the positions observed but JMG~99~G1 shows that the 
emission is dominated by relatively narrow lines 
(mostly between 0.4 and 1.5~\kms), typical of quiescent emission,
although line width can vary from region to region or between different species
(see on-line Tables~\ref{tb:obshh7a}, \ref{tb:obshh34ea}, \ref{tb:obshh34cd},
\ref{tb:obshh1} and \ref{tb:obsngc}).  HH~1 shows the largest differences
between line widths for different species, ranging from
$\sim0.5$~\kms\ for DCO$^+$ to 2.4~\kms\ for \met.  These large
variations are also observed in the HH~2 molecular clumps
(\cite{girart02}). For HH~7--11 positions and JMG 99 G2 most line
widths are in the 0.8 to 1.4~\kms\ range. Interestingly, for these
positions the HCN \J{3}{2} line emission is significantly broader than
the averaged line width (roughly by a factor 2). \tco\ lines are in
most positions either clearly non-Gaussian or have several components,
which can be explained from contribution from lower density components
along the line of sight. In these cases, in order to estimate the
integrated line intensity, we selected the velocity range where the
higher density tracer showed up. Strong absorption is observed at the
center of the \hco\ \J{1}{0}\ line in the HH~7-11 and specially in the
NGC 2264G region, which is due to the presence of a low density
molecular gas in the foreground (\cite{girart00}).

Some of the observed positions show high velocity wings in the \hco\
\J{1}{0}\ line (the three positions in HH~7--11, HH~1, HH~34 E and JMG
99 G2). This is more clearly seen in HH~7-11 A (see
Fig.~\ref{fhh7ahh34e}), where other strong lines also show wings:
\tco, H$_2$CO, CS, CH$_3$OH.

The properties of the molecular emission toward JMG 99 G1 is quite
different from the other regions. Some of the molecular transitions
have emission at the systemic velocity of the cloud (e.g. \cdo\
\J{2}{1}, \hco\ \J{4}{3}, CS
\J{3}{2}) but the strongest emission arises from a broad, $\Delta v \simeq
5$~\kms, blueshifted component. This component is also observed in SiO (\cite{garay02}). All these features indicate that most of the molecular emission
in JMG~99~G1 comes from shocked outflow material (see Section 5.3).

\section{Analysis}

The excitation temperature and the optical depth for the \hco\ can be
derived from the line ratio of the different \hco\ and \htco\ observed
transitions. From this analysis we found that $T_{\rm ex}\simeq$6--8~K
and $\tau$(\hco 3--2)$\simeq$3--11. Therefore \hco\ beam averaged
column density is derived adopting an excitation temperature of 7~K
and correcting for the optical depth.  For species which also have a
high dipole moment (e.g. HCN, H$_2$CO) the column densities are
estimated assuming $LTE$ and adopting $T_{\rm ex}=7$~K but also
assuming that the emission is optically thin.  For the other species
we adopt $T_{\rm ex}=10$~K. Column densities of CO and isotopes are
corrected for optical depth.  For a non detection we report the
3--$\sigma$ upper limit. Table~\ref{tb:cd} shows the derived beam
averaged column densities for the quiescent emission (i.e. excluding
the high velocity component) for all the positions but JMG~99~G1. For
reference, we also include the column densities in the SO$_2$ clump
ahead of HH~2 taken from Girart et al. (2005). Note that Girart et
al. obtained column densities for several positions along the slice in
the SO$_2$ clump (see their Table A.2); we chose to present only one
of them (position I3, T$_{ex}$ = 13$\pm$ 2 K).  Most of the lines
observed toward JMG~99~G1 show emission that is more characteristic of
strongly perturbed gas.  In addition we used the LVG and RADEX codes
to carry out a more detailed analysis for the molecules with several
transitions observed (see Table 4).

For the high velocity emission toward HH~7-11 and JMG 1 we also
estimated the column densities assuming $LTE$ and adopting $T_{\rm
ex}= 30$~K. In these cases, we computed the column density in
different velocity intervals along the velocity range here the high
velocity is detected (see \S5.3).

\subsection{LVG and RADEX analysis}

In this section we estimate, to a first approximation, some of the physical and
chemical characteristics of most clumps, such as the density and the
temperature by means of large velocity gradient (LVG) and RADEX modelling.

We observed multitransition observations of \hco\ and \htco\ for
HH~7--11~A, HH~7--11~B and HH~7--11~X, HH~1, HH~34~A and HH~34~E. We
therefore carried out radiative transfer calculations using the LVG
model approximation.  The main goal of the modelling is to constrain
the range of densities and temperatures of the clumps, as well as the
\hco\ true column densities.  LVG modelling is performed varying three
parameters: the column density of \hco\ , the temperature, and the
volume density of the gas; one parameter is provided as an input, and
the aim is to obtain a set of LVG solutions in the other two
parameters plane. For \hco, we ran the LVG code with the \hco\ column
density as input parameter and we stopped when a set of reasonable
solutions was obtained.  To constrain the set of solutions we used all
the observed line intensity as well as the observed line temperature
ratios ([H$^{12}$CO$^+$~3--2]/[H$^{13}$CO$^+$~3--2] and
[H$^{13}$CO$^+$~2-1]/[H$^{13}$CO$^+$~3--2]).  To account for the
difference in beam sizes between the different transitions, we
multiplied our intensities and temperature ratios by the appropriate
filling factor (assuming that the size of the emission is the same for
all the transitions and that the emission has a Gaussian shape).
We initially used the HCO$^+$ size derived from previous observations
(see following subsections), although we also explored different
values around the reported values to check if the LVG fits improved.
The solutions were searched in the $T_{k}$--$n({\rm H_2})$ plane for
different \hco\ column densities ranging from 1$\times$10$^{13}$ to
1$\times$10$^{15}$\cmcd\ (the \hco\ beam averaged column densities for
the clumps where we perform the LVG analysis are in the
2\N{13}--2\N{14} \cmcd\ range).  The results are summarized in
Table~\ref{tb:lvgrad}.  The range of values of the derived densities and temperatures from
the \hco\ LVG analysis was used to perform the same analysis with CS
for HH~7--11 and HH 34 E, using as completely free parameters only the CS column
density and the size of its emission.

SO and C$_3$H$_2$ are the only other species for which, for some objects, we
observed multiple
transitions and collisional rates are available.  For the SO (HH
7--11~A and HH 7--11~B) and C$_3$H$_2$ (HH 7--11~A) observed transitions we
used RADEX, a non--LTE molecular radiative transfer code which assumes an
isothermal homogeneous medium (\cite{sch05}). To find the possible
set of solutions for the SO and C$_3$H$_2$ we used the three observed
transitions  as well as their line ratios with the line intensities corrected
by the filling factor.  The set of best solutions  are shown in
Table~\ref{tb:lvgrad}.

\begin{table*}
\caption{Best fits with LVG and RADEX for column densities, gas densities and
temperatures. a(b) stands for a$\times$10$^b$. }
\begin{tabular}{|c|cccc|}
\hline
\multicolumn{5}{|c|}{\hco}  \\
\hline
Object& Size & $N$(\hco) (cm$^{-2})$ & $n({\rm H_2})$ (cm$^{-3})$ & $T_k$(K) \\
\hline
HH~7--11 A &  20$''$ & 5(14)-1(15) & 8(4)-3(5) & 15-24 \\
HH~7--11 B & 20$''$ & 5(14)-1(15) & 9(4)-4(5) & 15-24 \\
HH~7--11 X & 20$''$ & 5(14)-1(15) & 3(5)-2(6) & 7-11 \\ 
HH~1 & 10$''$ & 5(14)-1(15) & 4(5)-1(6) & 8-11 \\
HH 34 A & 10-15$''$ & 5(14)-1(15) & 3--4(5) & 7-11 \\
HH 34 E & 10-12$''$ & 5(14)-1(15) & 1.5-9(5) & 8--12 \\
\hline
\multicolumn{5}{|c|}{CS} \\
\hline
Object&Size &$N$(CS) (cm$^{-2})$ & $n({\rm H_2})$ (cm$^{-3})$ &  $T_k$(K) \\
\hline
HH~7--11 A & 30$''$ & 5(14)-1(15) & 1(5)-3(5)  &  15 \\
HH 34 E & 30$''$ & 3(13)--1(14)  & 3(5)-9(5)  &  10-12 \\

\hline
\multicolumn{5}{|c|}{SO} \\
\hline
Object & Size & $N$(SO) (cm$^{-2}$)         &$n({\rm H_2})$ (\cmt) &  $T_k$(K) \\
\hline 
HH 7--11 A  & 20$''$ & 3(14)-1(15)      & 3(4)-4(5)         & $>$10\\
HH 7--11 X  & 20$''$ & 3(14)-1(15)      & 2(4)-2(5)    & $>$10\\
\hline  
Object & Size & $N$(C$_3$H$_2$) (cm$^{-2}$) & $n({\rm H_2})$ (\cmt) &   $T_k$(K) \\
\hline  
HH 7-11 A  & 20$''$ & 3(13)-1(14)      & 4(4)-2(5)    & $>$10\\
     \hline
     \end{tabular}
\label{tb:lvgrad}
    \end{table*}
%

\subsubsection{HH~7--11}

\begin{figure}
\resizebox{7.5cm}{!}{\includegraphics{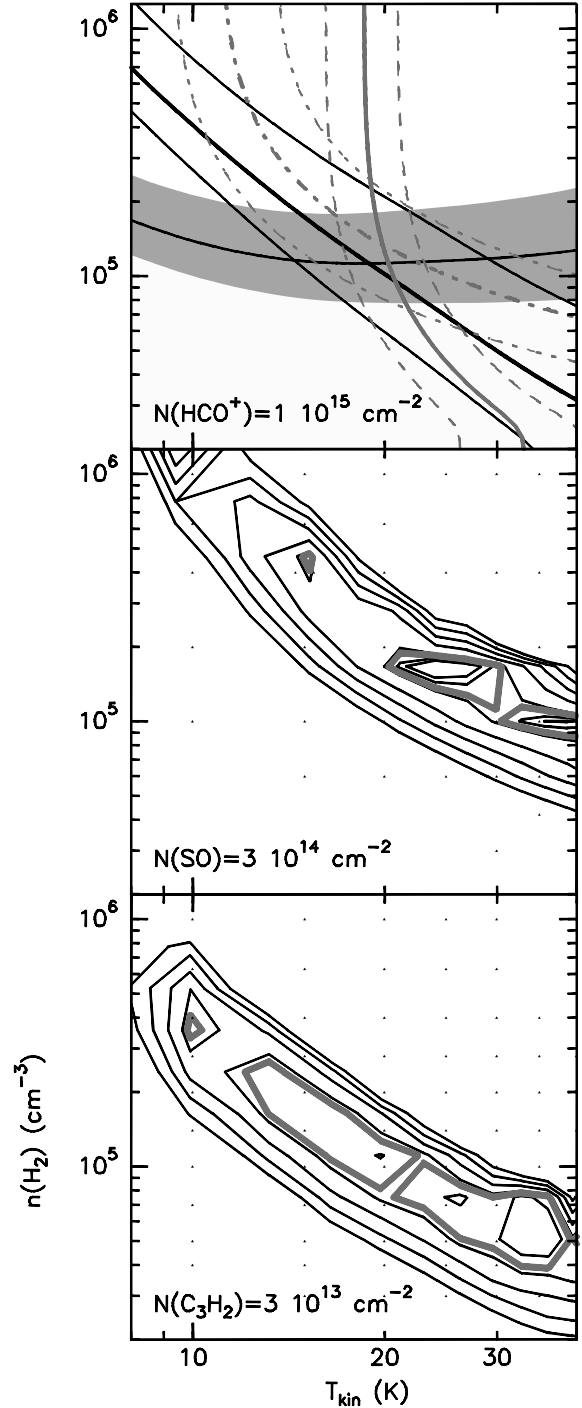}}
\hfill
\caption[]{
Plot of the set of LVG and RADEX solutions in the $T_{\rm K}$-$n$(H$_2$) plane
for HH 7--11~A assuming a source size of 20$''$ for the following molecules: 
{\it Top panel}  HCO$^+$, derived using LVG for a column density of
$10^{15}$\cmd. The grey scale and the thick solid line in its center shows the
\hco\ \J{3}{2}\ to \htco\ \J{3}{2}\ line ratio. The three solid lines show the
\htco\ \J{1}{0}\ to \htco\ \J{3}{2}\ line ratio. The thick solid and dashed
grey line show the  the  \hco\ \J{3}{2}\ and \hco\ \J{3}{2}\ lines,
respectively.
{\it Central panel}  SO, derived using RADEX for a column density of
3$\times10^{14}$ ~\cmd. The thin contours levels are (from lowest to highest
contour) $\chi^2 = 20,16,12,8,4,2,1$. The thick grey contour shows the 68\%
confidence region of $\chi^2$.
{\it Bottom panel}. C$_3$H$_2$, derived using RADEX for a column density of
3$\times10^{13}$ ~\cmd. The thin contours levels are (from lowest to highest
contour) $\chi^2 = 20,16,12,8,4$. The thick grey contour shows the 68\%
confidence region of $\chi^2$.
}
\label{fg:hcohh7a}
\end{figure}

For all the clumps we find good fits for \hco\ column densities
ranging from 5\N{14} to 1\N{15} \cmcd. As an example of the LVG
modelling, Figure~\ref{fg:hcohh7a} (top) shows a set of LVG solutions
in the $T_k$-$n({\rm H_2})$ plane for HH~7--11~A. The best fit is
obtained for a gas density of $\sim$ 8$\times$10$^4$--3$\times$10$^5$
cm$^{-3}$, a $T_k$ of 15--24~K and a source size of $\sim 20''$; while
for HH 7--11~B we find similar solutions, for HH~7--11~X we obtain a
higher gas density range and definitely a lower kinetic temperature.
Although HH~7--11~X seems close to HH~8, the low temperature may
be an indication that this clump is further away from the highest
excitation HH object. Also note that we do not know the 
line-of-sight component of the distance between clump
and HH object. The derived source size is larger than the
estimated size from Rudolph \& Welch (1988) maps. This may be due to
the clumps being very close to each other causing their emission to be
`contaminated', yielding to an apparently bigger size.

There are two CS transitions detected  toward HH~7--11~A, so the LVG analysis
have been performed by using as a constrain the lower limit of the  kinetic
temperature derived  from the LVG HCO$^+$ analysis, $T_{k}=$15~K.
Interestingly, we do not find any solution for a emitting size less than
30$''$,  suggesting  that the CS emission is more extended than the \hco\
one.   Adopting a size of 30$''$, the set of possible solutions spread out over
a large  range of volume and column densities. Table~\ref{tb:lvgrad} shows the
CS column density assuming a volume density similar to the value derived from
the \hco.

For HH~7--11~A and X we also performed RADEX analysis of SO and, for HH~7--11~
A, of C$_3$H$_2$. As for the other molecules we find that the best SO column
density fits are reached for values higher than the beam averaged ones. From SO
and C$_3$H$_2$ we were not able to constrain the temperature of the gas but we
notice that it has to be higher than 10~K (see Figure~\ref{fg:hcohh7a}, central
and bottom panel). The derived gas densities are slightly lower than those
derived from the \hco\ analysis.

\subsubsection{HH 1}

We did not detect the \htco\ 3--2 line, hence we had to adopt a
3--$\sigma$ upper limit for it as well as for the [H$^{12}$CO$^+ \,
3-2$]/[H$^{13}$CO$^+ \, 3-2$] and [H$^{13}$CO$^+ \,
1-2$]/[H$^{13}$CO$^+ \,3-2$] line ratios.  We find that, unlike for
HH~7--11 clumps, the best fits are obtained for a source size of
10$''$ consistent with the size found in the literature
(\cite{torrelles94}). This may support the idea that the apparent
larger size of the HH~7--11 clumps is merely a consequence of
contamination. While the \hco\ column and gas density of the HH~1
clump are similar to those for HH~7--11, its temperature is certainly
lower and somewhat better constrained.

\subsubsection{HH 34}

We find a remarkably small range of solutions for the kinetic
temperature, at 7-11~K for HH 34A and 8--11~K for HH 34 E. As for
HH~1, we find that the best solutions are obtained for 10$''$--15$''$
sizes, consistent with those found in the literature
(\cite{rudolph92}).  The \hco\ column densities range from 5\N{14} to
1\N{15} \cmcd. For HH 34A we derive slightly higher gas densities than
for HH 34 E.  For HH 34 E we also have the two CS transitions, hence
we attempted an LVG analysis by fixing the temperature to 10 and
15~K. As in the case of HH~7--11 there are no solutions for emitting
sizes smaller than $\sim30''$ and the CS column density is derived
assuming a volume density similar to the value derived from the \hco.

\subsubsection{Summary of LVG analysis}

In summary, the LVG analysis leads to the following preliminary
conclusions: $(i)$ The gas densities of the clumps are always higher
than those typical of a molecular cloud ($\sim$ 10$^4$ \cmgd); $(ii)$
\hco\ (and CS for HH~7--11 and HH 34 E) column densities derived from
LVG are $always$ larger than the beam averaged ones, consistent with
the fact that the size of the clumps is smaller than the beam sizes,
although in the case of the HH~7--11 clumps the picture is somewhat
more complicated due to the close vicinity of the clumps; $(iii)$ The
emitting sizes derived from the LVG analysis of CS seem to be larger
than those derived from the \hco\ analysis, indicating that the two
species possibly arise from different gas components, a result already
found by Girart et al. (2005) in their extensive study of the
molecular gas ahead of HH~2. $(iv)$ The temperatures are always well
constrained and for all the clumps lower than 25~K, with HH7-11 being
the warmest. Note that at low temperatures there is no clear pathway
for the enhancement of HCO$^+$: we show in VW99 that for HCO$^+$ to
reach column densities of $\ge$ 10$^{14}$ \cmcd\ grain surface
reactions and subsequent evaporation of the icy mantles must have
occurred, as the main route of formation for \hco\ is via the
ion-neutral reaction of C$^+$ with H$_2$O. If the temperature is $\le$
100~K, mantle evaporation can only have occurred via
photoevaporation. An alternative way might have been if the gas had
been shocked in the past reaching higher temperatures and then cooled
down. However during the time the gas cools back down to $\sim$
10--20~K, HCO$^+$ should be destroyed again. Moreover, we show in VW99
that an enhanced UV radiation is in any case needed in order to ionize
the gas enough to produce large abundances of C$^+$.

\section{Discussion}

As explained in Section 2,
our main criterion for extending the survey to other molecular species
was the strength of HCO$^+$. Table~\ref{tb:cd} shows that most clumps
that exhibit \hco enhancement also show a rich chemistry, as predicted
by the VW99: methanol, when observed, seems to be at least one order
of magnitude more abundant than what typically is found in dark
molecular clouds or prestellar cores (see Table~5). Of the sulphur
species, CS is ubiquitously detected but note however that CS is
abundant in dark clouds and hence some of the emission may arise from
the surrounding medium rather than the clumps, overall for those
clumps with a size much smaller than our beam size. SO and H$_2$S seem
to be present in the majority of the clumps where HCO$^+$ is
enhanced. Of the other sulphur bearing species, OCS, H$_2$CS, and NS
were only observed in HH~7--11~A. We will now
analyse our results and look for a possible correlation between the
chemical `richness' of a clump, its vicinity to the HH shock and the
strength of the radiation field impinging on the clump.

In this section we attempt a simple analysis of the molecular species detected
in our sample. In 5.1 we summarize the general physical and chemical properties
of all the clumps. In 5.2 we discuss the effects of distance on the chemical
processing ahead of HH objects by looking at the chemistry of different clumps
ahead of the same object. However, it is important to note that the abundance
of each molecular species will depend on  $i)$ the gas density; $ii)$ the
radiation field ; $iii)$ the distance to the source of radiation; $iv)$ the
size of the clump, $v)$ the age and, to a lesser extent, $vi)$ the kinetic
temperature of the gas.  Some of these physical parameters have similar effects
on the chemistry (e.g. the size and the gas density).
Due to such degeneracy it is often difficult to disentangle the effects on the
chemistry due to the different parameters.  So, for example, for clumps ahead
of the same object, where the radiation field is constant, different abundances
may not be due solely to the different positioning of the clumps as it may be
that the condensations differ in density, size and age. In 5.3 we briefly
discuss the properties of the high velocity gas.

\subsection{Physical and chemical properties of clumps ahead of HH objects}

The physical and chemical properties of the gas ahead of the HH
objects are different from those observed in prestellar cores or dark
molecular clouds: they are denser than the latter but their small size
and the high abundances of certain species, such as \hco\ and
methanol, make them unlikely candidates for pre-stellar cores. In
Table~\ref{tb:frac}, we compare the typical abundances of a
pre-stellar core, L1544, with those derived for the HH~7--11~A clump.
The abundances of L1544 are taken from J{\o}rgensen et al. (2002),
while the HH~7--11~A abundances were derived by dividing the averaged
beam column densities by the H$_2$ column density from the dust
observations (\cite{hatch05}) by assuming a dust temperature of 23~K
(\cite{chini01}). There are of course uncertainties associated
with the derivation of the H$_2$ column densities (and hence of
fractional abundances), such as i) the error bars associated with the
integrated emission (see Table 10) and ii) the assumption of
LTE which affects the molecular abundances estimate. 
While the first one only leads to a 30\% uncertainty, assuming
LTE may lead to a larger uncertainty. The latter is still
below the enhancement factor found (i.e $\sim$1 order
of magnitude), although,
for some species,
the propagation of both errors may become relevant. Although the abundances ahead of HH 7--11~A are
clearly enhanced with respect to those in L1544, the low abundance of
CO is puzzling: while freeze out on the grains is clearly responsible
for its depletion in L1544, in the case of HH~7--11~A the UV radiation
should have photoevaporated all the ices so no significant amount of
CO should be left on the dust unless it remains trapped in a
multilayered ice.  An alternative possibility is that the
condensations we observe are transient and young: if the UV radiation
photoevaporated the mantle while depletion was still occurring, there
may have been not enough time to form CO. Viti et al. (2003) already
observed that, in fact, to match the observations ahead of HH 2 the
condensations had to be young ($\le$ 1000 yrs) and in one of the best
matching models, B10, CO has a gas phase abundance of
2$\times$10$^{-5}$ at an A$_V$ of 3.5 mags after evaporation.  On the
other hand it is also possible that the dust column density is
overestimated: some of the material may not be associated with the
dense clumps and we may be overestimating the amount of dust
responsible for the emission, perhaps because we have a too low
an opacity. In short, it is hard to know what the absolute abundance of
CO is and no strong conclusion should be drawn from
Table~\ref{tb:frac} beside the fact that clearly some species are
enhanced with respect to those in L1544.

An enhancement of species such as
\hco, \met\  and SO, in absence of shocks, can indeed be an indication of
active photochemistry.  J{\o}rgensen et al. (2002) do not give a
fractional abundance for
\met\  for L1544 but \hco\ and SO are both less abundant by a factor of 5--10,
although species such as CS have similar abundances.  Interestingly
DCO$^+$ also seems to be slightly enhanced in HH~7--11~A with
respect to the value found in L1544, although the DCO$^+$/\hco\ ratio
is consistent with those expected in a dense, low temperature gas
(\cite{roberts00}). In fact a comparison between starless cores and
clumps ahead of HH objects may be misleading in that the anomaly of
our clumps is that they are denser than a molecular cloud but much
smaller than starless cores. They in fact show similarity with clumps
detected by recent high resolution BIMA observations in L673
(\cite{morata05}) in that they are of similar sizes and they seem to
have structures on scales of $\sim$ 10$''$ or so but they have
different chemistry because they are illuminated by the HH radiation.

\begin{figure}
\vspace{7cm}
\includegraphics{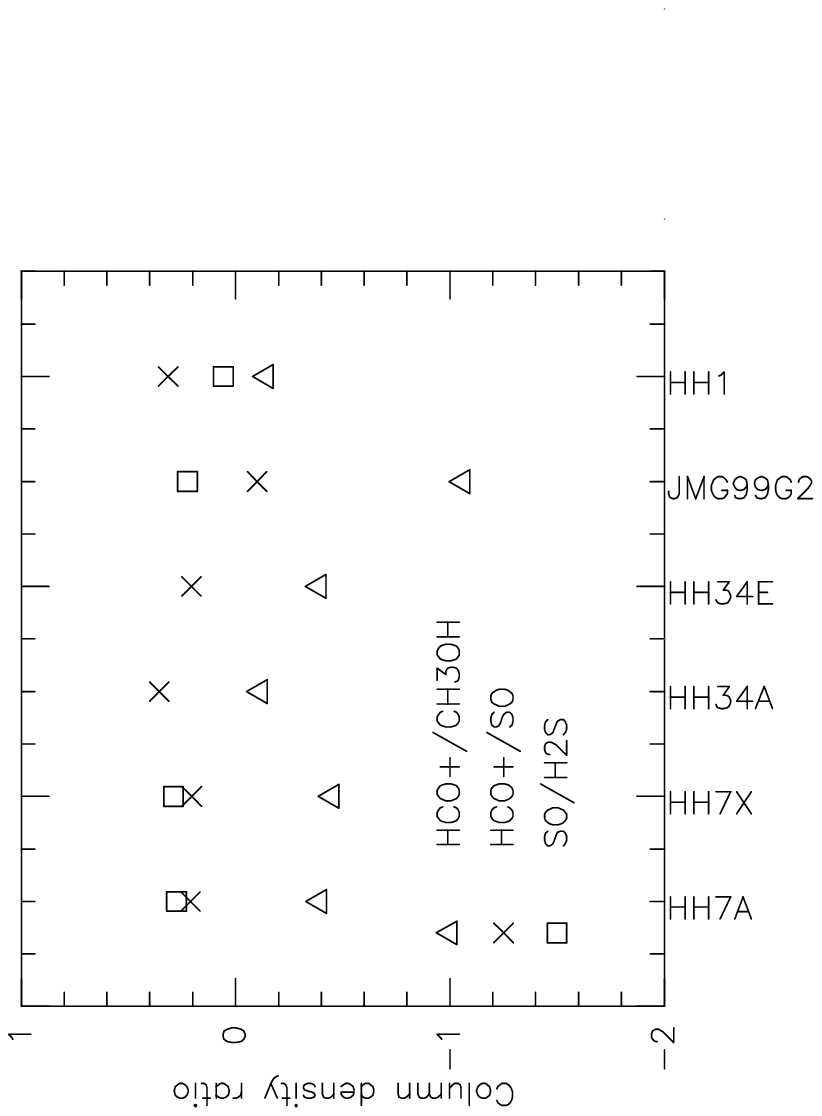}
\caption{Ratios (in log 10 scale) of the column densities of selected species for the low velocity gas. HH~7 B, HH~34 C, HH~34 D and
JMG 99 G1 are not shown due to lack of data.}
\label{fg:ratios}
\end{figure}

We are unable to compare clumps ahead of other regions with starless cores as
we do not know their H$_2$ column density but we compare the column
densities across our clumps in Table~\ref{tb:cd} and Figure~\ref{fg:ratios}.
\begin{table}
\caption{Fractional abundances with respect to H$_2$ for HH~7-11 and L1544}
\begin{tabular}{|c|cc|c|}
\hline
\multicolumn{1}{|c|}{} &
\multicolumn{2}{|c|}{HH~7--11} & \\
Molecule & A & X & L1544 \\
\hline
CO     & 3.7(-5) & 5.5(-5) & 4.9(-6) \\
\met\  & 5.0(-9) & 9.8(-9) & --      \\
\hco\  & 2.9(-9) & 3.8(-9) & 3.9(-10)\\
H$_2$CO& 1.3(-9) & 2.3(-9) & --      \\
SO     & 1.2(-9) & 2.0(-9) & 4.8(-10)\\
CS     & 8.6(-10)& 1.5(-9) & 8.6(-10)\\
H$_2$S & 7.1(-10)& 1.1(-9) & --      \\
HCN    & 3.3(-10)& 3.8(-10)&$<$4(-10)\\
OCS    & 3.0(-10)& --      & --      \\
H$_2$CS& 1.6(-10)& --      & --      \\
DCO$^+$& 7.1(-11)& 8.6(-11)& 2.1(-11)\\
\hline
\end{tabular}
\label{tb:frac}
\end{table}

Before attempting such comparison, it is important to note that, as
discussed in Section 3 and as shown from our LVG analysis, we are
uncertain about the size of all our clumps; hence for some objects,
the beam size may be larger than the source size and for species such
as HCO$^+$ (enhanced only in dense clumps subjected to the radiation)
their averaged beam column densities may be a lower limit; however for
species that are abundant also in the surrounding medium, such as CS
and CO, their beam averaged column densities may be an overestimate of
the emission from the clump.

\par 
From Table~\ref{tb:cd} it is clear that all the clumps, except
possibly JMG 99 G2, exhibit a similar chemistry: in order to eliminate
the uncertainties due to the sizes, let us compare ratios rather than
absolute column densities (see Figure 7); for example, \hco/\met\ is remarkably
constant within a factor of 2 for all objects, except for G2, where it
is lower by one order of magnitude. Other ratios, such as
\hco/SO or SO/H$_2$S also show remarkable consistency among the clumps
although in G2 the column densities of species characteristic
of photochemistry are not as enhanced. In fact, chemically, it appears
that while the clumps ahead of HH~7--11, HH 34 and HH~1 are of the
same nature (confirmed also by the column densities derived from the
LVG analysis), JMG 99 G1 and G2 have undergone different processing.
\par
Although we do not have an estimate for the temperature of JMG 99 G1
and G2, from our LVG analysis we note that HH~7--11~A clumps have a
higher temperature than HH 34 and HH~1 clumps, therefore HH 34 and
HH~1 clumps seem to be the most quiescent, maybe because so far they
only have been affected by the UV radiation. Indeed, the line widths
of HH~34 are significantly smaller than those of HH~7-11 ($\Delta v
({\rm H^{13}CO^+})\simeq 0.4$ and 1.2~km~s$^{-1}$ for HH~34 and
HH~7--11, respectively).  HH~7--11 clumps, on the other hand, show
sign of dynamical interaction with the HH objects, but such
interaction seems not to have affected the chemical nature of the gas
yet: the increase in temperature caused by the dynamical interaction
is in fact still not sufficient to produce any shock chemistry and
hence the chemical properties of HH~7--11 are very similar to those of
HH 34 and HH~1 clumps.
\par
Note that, as also explained in Viti et al. (2004), the density
enhancements we see ahead of HH~7--11 could not be produced by wind
compression as the timescales would be too short for substantial
freeze out to have occurred; as described in VW99 and Viti et
al. (2003) depletion and subsequent hydrogenation on the grains must
have occurred during the formation of the density enhancements, before
the arrival of the HH object, in order for species such as \hco\ and
\met\ to have formed.
\par
It is tempting to draw an evolutionary scenario around these three
different situations with HH 34 and HH~1 clumps being the youngest and
JMG 99 G1 and G2 being the oldest clumps. However, it is impossible to
disentangle the effects of age, distance from the HH objects, and
radiation, as discussed further in the next section.
\par
Despite the similar chemistry of most of the clumps and the
uncertainties with the emission sizes, the (albeit small) differences
in column densities (Table 3) may be significant.  The three clumps
ahead of HH~7--11 may be affected by several HH objects
(Figure~\ref{fhh7a}), hence it is impossible to determine which
clump is the most irradiated; however, irradiation from several HH
objects is consistent with the fact that HH~7--11 clumps show a richer
chemistry than others.
\par
Although we do not here attempt any modelling of these clumps, it is
interesting to compare the abundances of our clumps with the models
computed in Viti et al. (2003). Note that these were detailed models
of a particular clump ahead of HH~2 hence we are not looking here for
a perfect match. Viti et al. (2003) computed a large grid of models
where both the radiation field and densities were varied. Although
from the column densities in Table 3 it seems that HCO$^+$ is higher
ahead of HH~7--11 than ahead of HH 34 and NGC2264, from our LVG
analysis it seems that the densities ahead of HH~7--11 are either
lower or the same as those ahead of HH 34 and HH~1. Hence we may
expect either that the radiation field from HH~7--11 is higher, or
that clumps ahead of HH~7--11 are younger.  Figures 3 to 7 in Viti et
al. (2003) show the column densities for \hco, \met\ , SO and H$_2$CO
as a function of time for an A$_V$ $\sim$ 3 mags. We compare the
column densities in these figures with those derived from our
observations here for HH~7--11 and find that the best fitting models
for HCO$^+$, CH$_3$OH and SO are B7 and B8 which have a high radiation
field of 1000 Habing and a density of 3$\times$10$^5$ \cmgd.  The
abundances of \hco, \met\ and SO may be less ahead of HH 34 by as much
as 1 order of magnitude: although not shown in Viti et al. (2003) we
compared two models from their grid where the density (3\N{5} \cmgd),
age ($\le$ 100 yrs) and visual extinctions ($\sim$ 3 mags) were the
same but the radiation field varied from 20 to 1000 Habing and we
found that \hco went from 7\N{12} to 5\N{13} \cmcd, SO from 5\N{14} to
2\N{15} and \met\ from 1\N{14} to 6\N{14} \cmcd. However, at much
later ages, a higher radiation field implies in fact a lower
abundances of these species as the radiation field enhances but also
destroys the gas: as explained in VW99 the effect of the irradiation
is in fact to create a `wave' of molecular formation where the
abundances rise to enhanced values with respect to standard dark cloud
chemistry but then fall again as photodissociation begins to
dominate. It is difficult therefore to disentangle the effects of
radiation, from those of age.
\par
Finally, it is interesting to note that in HH~7--11~A, the abundance of
HCS$^+$ is at least one order of magnitude lower than that of H$_2$CS.
Again, from the \cite{viti03} models we find that, as long as the
density is greater than 10$^5$ \cmgd then H$_2$CS/HCS$^+$ is always
$\ge$ 10; but lower radiation fields ($\sim$ 20 Habing) yield to very
high ratios (100-1000) while if $\chi$ $\sim$ 1000 Habing the ratio is
$\sim$ 10-50.

\subsection{Effects of distance on chemical processing ahead of HH objects}

In this section we attempt to single out the effect of the distance
from other parameters under the assumption that the rich chemistry is
mainly due to the UV impinging on the clumps. We therefore compare
column densities among clumps belonging to the $same$ region, where
not only the radiation field from the HH shock is constant, but also
where we assumed that the size of each clump is the same. 
In particular HH 34 may be the best
region for this analysis as i) the four clumps show some chemical
diversity and ii) unlike the HH~7--11 region there seems to be only
one HH object impinging on the clumps.
\par

In the HH~34 region HH~34~C and E are the closest clumps to the
HH~object, with A and D 1.6 and 2 times further away (in the plane
of the sky; the true distances from the HH object are of course
unknown).  From Table~\ref{tb:cd}, HH~34~C and E clearly show higher
\hco\ column densities than A and D, although the CO and CS column
densities are similar, consistent with exposure to a higher UV
field.  Of C and E, E has the larger \hco\ column density, possibly
because it has the higher derived gas density derived from the LVG
analysis. Moreover, from Table~\ref{tb:cd}, SO and methanol from HH
34~E have higher column densities than the more distant HH~34~A.
From this object, we also detect C$_3$H$_2$, not usually detected in
quiescent dense clumps.  This increase in chemical activity where
the UV radiation field would be expected to be strongest is
consistent with the UV chemistry, though a variation in density or
age of the clumps cannot be ruled out as an alternative cause.

\subsection{The High Velocity component}
\begin{figure}
  \resizebox{9.0cm}{!}{\includegraphics{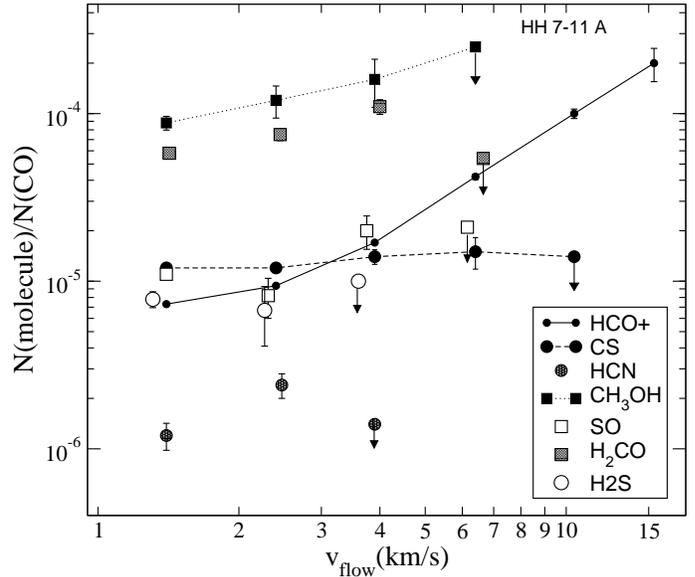}}
     \caption[]{
 Fractional abundance with respect to the CO of the high velocity gas for
several molecules toward HH 7--11 A.
$v_{\rm flow} = v_{\rm LSR} - v_{\rm amb}$, where
$v_{\rm amb}$ is the systemic velocity of the ambient gas. }
\label{fhighvel}
\end{figure}

HH~7--11 A and JMG 99 G1 are the positions where the high velocity
emission is brighter. We calculated the column density ratio
  with CO for several species (from the \tco\ spectra), i.e.,
the molecular fractional abundances with respect to the CO 
(see \S4).
Figure~\ref{fhighvel} shows the fractional abundances with respect to
the CO for several molecules as a function of the outflow velocity.

From Figure 8, it is clear that the \hco\ relative abundance of the HH~7--11 high velocity
component toward increases by more than an order of magnitude (a factor
of $\sim 30$) between the slowest and fastest outflow components. This
enhancement is also seen in H$_{2}$CO and
\met\, for outflow velocities lower than $\sim 6$~\kms\ with respect to
the systemic velocity.  On the other hand, CS does not change its
relative abundance.  The \hco\ abundance enhancement at higher outflow
velocities was first reported by Girart et al. (1999) towards the NGC
2071 molecular outflow. Recently, Dent et al. (2003) has found the same
behaviour for the \hco\ towards the high velocity gas around
HH~2. However, HCO$^+$ is the only high density molecular tracer
detected in the high velocity HH~2 component and this case could be a
different scenario from NGC~2071 or HH~7--11, since in HH~2 the lack of
other molecular high velocity gas other than HCO$^+$ can be explained
by the presence of strong UV field. In any case the clear enhancement
of the HCO$^+$ abundance with respect to other molecular tracers with
increasing high velocity indicates that it traces the molecular
outflow well at higher velocities, whatever causes
its enhancement.

JMG 99 G1 not only has a spectral feature quite different from the
high velocity wings seen in HH~7--11 A but the chemical properties are
also very different. In JMG 99 G1 the \met\ is by far the more
abundant high density molecular tracer with a 
X[\met]/X[CO]$\simeq$2--4$\times10^{-3}$. 
This abundance is also
found in strongly shock molecular clumps such those in L1157
(\cite{bach01}), and IRAS 21391+5802 (\cite{beltr02}). The \hco on the other hand
is almost one order of magnitude lower than in HH~7--11, although its
abundance also increases with the outflow velocity. Even the
spectral shape of the molecular emission in G1 is similar to the
spectral shape of the lines detected in the L1157 clumps:
Most of the lines shown are blueshifted (by 2--3~km~s$^{-1}$ with respect to the
cloud velocity) and are significantly broader ($\Delta v \simeq$
4--5 km~s$^{-1}$) than the lines from the other positions. Such
emission may in fact come from clumps similar to those seen along the
L1157 and CB3 outflows (\cite{bach01}; \cite{codella99}): Viti et
al. (2004) found that such clumpiness may be a consequence of a
pre-existing density enhancement and of its interaction with the
outflow.

\section{Conclusions}
We have performed a survey of molecular species ahead of several HH
objects, where enhanced \hco\ emission was previously observed. Recent
modelling have supported the idea that such enhancement is due to UV
radiation from the HH object lifting the ice mantles from the grains
in the clumpy medium ahead of the jet, which then would drive a
photochemistry. If this model is correct then other molecular species
(such as \met, SO etc) should also be enhanced. A recent molecular
survey of the gas ahead of HH~2 seems to support such model
(\cite{girart05}). The aims of this present survey were to
investigate: 1) whether the rich chemistry observed ahead of HH~2, and
predicted by the VW99 model, was indeed characteristic of the gas
ahead of $all$ Herbig-Haro objects; 2) the chemical differentiation
that may arise among clumps ahead of different objects, which may
differ in radiation field strengths.
\par
We have selected 4 HH objects regions: in some of these regions
enhanced \hco\ emission had been detected from several clumps. Hence, in
total we observed ten positions.  For each position we searched for
more than one species and, if possible, for more than one transition
per species. We find that, chemically, the richest gas was that ahead
of the HH 7--11 complex, in particular at the HH~7--11 A position. In
some regions we also detected a high velocity gas component. Beside
estimating the averaged beam column densities for all the species
observed, we have performed, where possible, LVG analyses to constrain
the physical properties of the gas. The main conclusions from our LVG
analysis are:
\begin{enumerate}
\item The gas densities of all the clumps observed are always higher than those
typical of molecular clouds by at least a factor of 5, but probably higher.
\item The derived gas temperatures are alway quite low ($<$ 15 K) apart for the
clumps ahead of HH~7--11 where the maximum temperature does not in any case
exceed 25 K.  
\item \hco, SO and CS column densities derived from our LVG and RADEX analyses
are higher than the beam averaged ones. This indicates that the size of the
emitting region is smaller than the beam size.
\item The emitting size derived from the LVG analysis of CS are larger than
those estimated from the \hco\ analysis. It may be that these two
species arise from different gas components, consistent with previous
studies (\cite{girart05}). In addition, the \hco\ abundance is
enhanced with respect to the starless core in L1544, whereas the CS
has similar abundances. All this suggests that at least a fraction of
the CS emission may arise from a cloud component or clumps not
illuminated by he HH objects.
\end{enumerate}

By comparing column densities among the several clumps we also attempt
a very simple chemical analysis in light of the photochemical model.
We find that the abundances of the observed species ahead of the
surveyed HH objects are higher than those found in a typical starless
core and that by comparing the column densities among the different
clumps we find that the clumps ahead of HH~7--11, HH~34 and HH~1 are
of the same chemical nature, supporting the idea that the gas
ahead of all HH objects exhibit a rich chemistry.
\begin{acknowledgements}
SV and JH both acknowledge individual financial support from a PPARC Advanced Fellowship. JMG acknowledges support by SEUI grant AYA2002-00205.
SV and JMG acknowledge support by a joined RS and CSIC travel grant.  
We thank the staff of the IRAM 30 m telescope for help with the
observations.
\end{acknowledgements}

\clearpage
\begin{sidewaystable*}
\begin{minipage}[180mm]{\textwidth}
\caption{Molecular line results for HH 7--11}
\label{tb:obshh7a}
\centering
\renewcommand{\footnoterule}{}  
\begin{tabular}{lcccccccccccccc}
     \noalign{\smallskip}
          \hline
	       \noalign{\smallskip}   
	       \multicolumn{1}{c}{} &
	       \multicolumn{5}{c}{HH 7--11 A} &
	       \multicolumn{5}{c}{HH 7--11 X} &
	       \multicolumn{4}{c}{HH 7--11 B} 
	       \\
	       \cline{2-5}\cline{7-10}\cline{12-15}
	       \multicolumn{1}{l}{Molecular} &
	       \multicolumn{1}{c}{$T_{\rm MB}$} &
	       \multicolumn{1}{c}{$\!\!\!\int T_{\rm MB} dv$} & 
	       \multicolumn{1}{c}{$v_{\rm LSR}$} &
	       \multicolumn{1}{c}{$\Delta v_{\rm LSR}$} &
	       \multicolumn{2}{c}{$T_{\rm MB}$} &
	       \multicolumn{1}{c}{$\!\!\!\int T_{\rm MB} dv$} & 
	       \multicolumn{1}{c}{$v_{\rm LSR}$} &
	       \multicolumn{1}{c}{$\Delta v_{\rm LSR}$} &
	       \multicolumn{2}{c}{$T_{\rm MB}$} &
	       \multicolumn{1}{c}{$\!\!\!\int T_{\rm MB} dv$} & 
	       \multicolumn{1}{c}{$v_{\rm LSR}$} &
	       \multicolumn{1}{c}{$\Delta v_{\rm LSR}$} 
	       \\
	       \multicolumn{1}{l}{Transition} &
	       \multicolumn{1}{c}{(K)} &
	       \multicolumn{1}{c}{$\!\!\!\!\!$(K\kms)} & 
	       \multicolumn{1}{c}{$\!\!\!\!\!$(km~s$^{-1}$)} &
	       \multicolumn{1}{c}{$\!\!\!\!\!$(km~s$^{-1}$)} &
	       \multicolumn{2}{c}{(K)} &
	       \multicolumn{1}{c}{$\!\!\!\!\!$(K\kms)} & 
	       \multicolumn{1}{c}{$\!\!\!\!\!$(km~s$^{-1}$)} &
	       \multicolumn{1}{c}{$\!\!\!\!\!$(km~s$^{-1}$)} &
	       \multicolumn{2}{c}{(K)} &
	       \multicolumn{1}{c}{$\!\!\!\!\!$(K\kms)} & 
	       \multicolumn{1}{c}{$\!\!\!\!\!$(km~s$^{-1}$)} &
	       \multicolumn{1}{c}{$\!\!\!\!\!$(km~s$^{-1}$)} 
	       \\
	            \noalign{\smallskip}     
		         \hline
			      \noalign{\smallskip}     
			      \tco\ \J{2}{1}$^a$        &15.8(2) &23.8(1)  & 7.8(1) & 1.5(2) &&18.6(3) & 26.3(1) & 7.7(1) & 1.7(2) && 16.3(4) &24.5(2) & 7.8(1) & 1.4(2)\\
			      \cdo\ \J{2}{1}$^a$        & 4.86(6)& 6.9(1)  & 7.8(1) & 1.4(2) && 4.8(2) &  6.76(7)& 7.8(1) & 1.5(2) && & & & \\
			      \hco\ \J{1}{0}$^a$        & 6.51(8)&10.37(3) & 7.80(5)& 1.69(8)&& 6.33(14)&11.51(6)& 7.8(1) & 1.4(2) && 6.36(14)&10.0(1) & 7.9(1) & 1.4(2)\\
			      \hco\ \J{3}{2}            & 9.5(3) &18.08(18)& 7.87(1)& 1.80(2)&& 8.5(6) & 17.7(4 )& 7.92(2)& 1.94(5)&&10.0(6) & 18.1(4) & 7.8(1) & 1.6(2)\\
			      \htco\ \J{1}{0}           & 1.91(3)& 2.43(2) & 7.92(1)& 1.20(1)&& 1.71(5)&  2.21(3)& 7.89(1)& 1.21(2)&& 1.92(6)&  2.44(4)& 7.97(1)& 1.20(2)\\
			      \htco\ \J{3}{2}           & 1.0(2) & 1.08(8) & 8.03(4)& 1.03(9)&& 0.71(18)& 0.95(9)& 8.11(6)& 1.26(12)&&1.1(3) &  1.3(2) & 8.21(8)& 1.2(2) \\
			      \dco\ \J{2}{1}            & 5.00(14)&5.42(6) & 7.86(2)& 1.02(1)&& 3.7(2) &  3.78(8)& 7.73(1)& 0.97(2)&& & & & \\
			      \dco\ \J{3}{2}            & 3.5(2) & 3.45(8) & 7.84(1)& 0.91(2)&& 2.1(3) &  2.48(15)&7.75(3)& 1.10(7)&& & & & \\
			      HCN \J{3}{2}              & 2.2(4) & 7.25(31)& 7.82(7)& 3.2(2) && 2.0(4) &  5.1(5) & 8.00(9)& 2.4(3) && & & & \\
			      \for\ \JK{2}{0,2}{1}{0,1} & 4.2(2) & 7.06(7) & 7.68(1)& 1.59(2)&& 3.51(18)& 7.23(12)&7.71(2)& 1.93(4)&& & & & \\
			      \for\ \JK{3}{1,3}{2}{1,2} & 4.2(1) & 6.42(3) & 7.86(1)& 1.42(1)&& & & & && & & & \\
			      \met\ \JK{2}{0}{1}{0} A$^+$&2.62(5)& 3.60(3) & 7.83(1)& 1.29(1)&& 2.38(7)&  3.53(4)& 7.84(1)& 1.40(1)&& & & & \\
			      \met\ \JK{2}{-1}{1}{-1} E & 1.93(5)& 2.67(3) & fixed  & fixed  && 1.74(7)&  2.60(4)& fixed  & fixed  && & & & \\
			      \met\ \JK{2}{0}{1}{0} E   & 0.36(5)& 0.49(3) & fixed  & fixed  && 0.29(7)&  0.44(4)& fixed  & fixed  && & & & \\
			      \met\ \JK{2}{1}{1}{1} E   & 0.11(5)& 0.16(3) & fixed  & fixed  && 0.09(7)&  0.12(4)& fixed  & fixed  && & & & \\
			      \met\ \JK{3}{0}{2}{0} A$^+$&3.13(10)&4.33(7) & 7.86(1)& 1.30(1)&& 2.91(11)& 5.08(9)& 7.96(1)& 1.59(2)&& & & & \\
			      \met\ \JK{3}{-1}{2}{-1} E & 2.53(10)&3.50(7) & fixed  & fixed  && 2.59(11)& 4.54(9)& fixed  & fixed  && & & & \\
			      \met\ \JK{3}{0}{2}{0} E   & 0.25(10)&0.35(6) & fixed  & fixed  && 0.53(11)& 0.99(8)& fixed  & fixed  && & & & \\
			      \met\ \JK{5}{0}{4}{0} A$^+$&1.01(16)&1.42(14)& 7.82(5)& 1.33(7)&& 1.5(3) &  2.3(2) & 8.02(6)& 1.39(9)&& & & & \\
			      \met\ \JK{5}{-1}{4}{-1} E & 0.96(16)&1.35(12)& fixed  & fixed  && 1.3(3) &  1.8(2) & fixed  & fixed  && & & & \\
			      \cthc\ \JK{5}{0}{4}{0}    & 0.14(4)& 0.13(1) & 7.87(3)& 0.85(4)&& & & & && & & & \\
			      \cthc\ \JK{5}{1}{4}{1}    & 0.14(4)& 0.13(1) & fixed  & fixed  && & & & && & & & \\
			      \cthc\ \JK{5}{2}{4}{2}    & 0.08(4)& 0.07(1) & fixed  & fixed  && & & & && & & & \\
			      \cthd\ \JK{2}{1,2}{1}{0,1}& 1.43(12)&1.33(3) & 8.10(1)& 0.86(2)&& & & & && & & & \\
			      \cthd\ \JK{3}{1,2}{2}{2,1}& 0.53(10)&0.69(8) & 7.87(1)& 1.21(10)&&0.50(11)& 0.65(8) & 7.96(1)& 1.21(11)&& & & & \\
			      \cthd\ \JK{4}{1,4}{3}{0,3}& 1.17(11)& 0.91(4)& 8.10(2)& 0.73(4)&& & & & && & & & \\
			      CS \J{3}{2}               & 6.5(2) & 12.10(9)& 7.73(1)& 1.72(2)&& 6.2(3) & 12.6(2) & 7.70(2)& 1.93(4)&& 6.3(3)& 12.0(2) & 7.84(1)& 1.77(1)\\
			      CS \J{5}{4}               & 2.9(1) &  4.00(4)& 7.90(1)& 1.26(1)&& & & & && & & & \\
			      SO \JK{3}{2}{2}{1}        & 5.43(9)&  7.23(4)& 7.80(1)& 1.25(1)&& 5.04(8)&  6.73(4)& 7.78(1)& 1.26(1)&& & & & \\
			      SO \JK{6}{5}{5}{4}        & 2.7(2) &  3.06(9)& 7.73(1)& 1.05(3)&& 2.0(2) &  3.09(12)&7.93(3)& 1.39(7)&& & & & \\
			      SO \JK{7}{6}{6}{5}        & 1.1(3) &  1.7(2) & 7.61(8)& 1.4(2) && 0.8(3) &  1.1(2) & 7.8(1) & 1.2(4) && & & & \\
			      H$_2$S \JK{1}{1,0}{0}{0,1}& 2.8(2) &  4.17(11)&7.73(2)& 1.41(4)&& 2.1(3) &  3.9(2) & 7.77(5)& 1.75(11)&& 2.2(3)& 3.5(2) & 7.92(5)& 1.50(9)\\
			      \htcs\ \JK{4}{0,4}{3}{0,3}& 0.49(9)&  0.48(5)& 7.61(5)& 0.90(9)&& & & & && & & & \\
			      NS \J{7/2}{5/2} 161.297   & 0.16(7)&  0.18(3)& 7.94(7)& 1.05(8)&& & & & && & & & \\
			      NS \J{7/2}{5/2} 161.298   & 0.14(7)&  0.15(7)& fixed  & fixed  && & & & && & & & \\
			      NS \J{7/2}{5/2} 161.302   & 0.14(7)&  0.16(3)& fixed  & fixed  && & & & && & & & \\
			      OCS \J{7}{6}              & 0.25(7)&  0.21(2)& 7.76(5)& 0.79(9)&& & & & && & & & \\
			      HCS$^+$ \J{5}{4}          &$\la0.21$&$\la0.10$& & && & & & && & & & \\
			           \noalign{\smallskip}
				   \hline
				   \end{tabular}
				        \begin{list}{}{}
					\item[$^a$] The $^{13}$CO 1-0, C$^{18}$O 1-0, HCO$^+$ 1-0 lines are clearly
					non-Gaussian (self absorption for the HCO$^+$ and strong high velocity emission
					for all of them). Line parameters derived from the zero (integrated
					intensity),  first (line velocity) and second (line width) order moments of
					the  emission in the 6.9-8.9 v$_{\rm LSR}$ range (the velocity range of the
					\htco\ emission). 
					     \end{list}
					     \vfill
					     \end{minipage}
					     \end{sidewaystable*}

\clearpage
%

     \begin{table*}
     \caption[]{Molecular line results for HH 34 E \& A}
     \label{tb:obshh34ea}
     \[
     \begin{tabular}{lcccc@{\hspace{0.1cm}}ccccc}
     \noalign{\smallskip}
     \hline
     \noalign{\smallskip}   
\multicolumn{1}{c}{} &
\multicolumn{5}{c}{HH 34 A} &
\multicolumn{4}{c}{HH 34 E} 
\\
\cline{2-5}\cline{7-10}
\multicolumn{1}{l}{Molecular} &
\multicolumn{1}{c}{$T_{\rm MB}$} &
\multicolumn{1}{c}{$\int T_{\rm MB} dv$} & 
\multicolumn{1}{c}{$v_{\rm LSR}$} &
\multicolumn{2}{c}{$\Delta v_{\rm LSR}$} &
\multicolumn{1}{c}{$T_{\rm MB}$} &
\multicolumn{1}{c}{$\int T_{\rm MB} dv$} & 
\multicolumn{1}{c}{$v_{\rm LSR}$} &
\multicolumn{1}{c}{$\Delta v_{\rm LSR}$} 
\\
\multicolumn{1}{l}{Transition} &
\multicolumn{1}{c}{(K)} &
\multicolumn{1}{c}{$\!\!\!\!\!$(K\kms)} & 
\multicolumn{1}{c}{$\!\!\!\!\!$(km~s$^{-1}$)} &
\multicolumn{2}{c}{$\!\!\!\!\!$(km~s$^{-1}$)} &
\multicolumn{1}{c}{(K)} &
\multicolumn{1}{c}{$\!\!\!\!\!$(K\kms)} & 
\multicolumn{1}{c}{$\!\!\!\!\!$(km~s$^{-1}$)} &
\multicolumn{1}{c}{$\!\!\!\!\!$(km~s$^{-1}$)} 
\\
     \noalign{\smallskip}     
     \hline
     \noalign{\smallskip}     
\tco\ \J{2}{1}$^a$        &11.7(3)  &10.04(2) & 8.60(9)& 0.65(9) &&12.2(3)  &11.15(8) & 8.7(1) & 0.6(1)  \\
\cdo\ \J{2}{1}$^a$        & 5.51(15)& 2.70(4) & 8.70(1)& 0.46(1) && 4.2(2)  & 2.65(8) & 8.73(1)& 0.64(2) \\
\hco\ \J{1}{0}            & 4.06(9) & 2.31(3) & 8.49(1)& 0.54(1) && 2.85(12)& 2.00(5) & 8.49(1)& 0.66(2) \\
\hco\ \J{3}{2}            & 2.3(4)  & 1.14(13)& 8.51(3)& 0.46(5) && 2.4(5)  & 1.69(19)& 8.63(4)& 0.65(10)\\
\htco\ \J{1}{0}           & 0.72(8) & 0.31(2) & 8.63(1)& 0.40(3) && 0.91(7) & 0.44(2) & 8.78(1)& 0.46(2) \\
\htco\ \J{3}{2}           & 0.6(2)  & 0.20(6) & 8.3(1) & 0.3(1)  &&$\la1.0$ &$\la0.22$&        &         \\
\dco\ \J{2}{1}            & 1.2(2)  & 0.46(3) & 8.54(1)& 0.37(3) && 1.25(15)& 0.55(3) & 8.69(1)& 0.42(2) \\
\dco\ \J{3}{2}            & 0.7(2)  & 0.26(4) & 8.55(4)& 0.37$^b$&& 0.67(16)& 0.20(4) & 8.43(5)& 0.3(1)  \\
HCN \J{3}{2}              &$\la0.75$&$\la0.20$&        &         &&$\la0.75$&$\la0.20$&        &         \\
\for\ \JK{2}{0,2}{1}{0,1} & 1.9(2)  & 0.76(5) & 8.48(1)& 0.38(4) && 1.5(3)  & 0.86(12)& 8.56(3)& 0.56(10)\\
\for\ \JK{3}{1,3}{2}{1,2} &         &         &        &         && 1.43(11)& 0.87(3) & 8.78(1)& 0.57(2) \\
\met\ \JK{2}{0}{1}{0} A$^+$&0.68(7) & 0.31(2) & 8.64(1)& 0.43(2) && 1.41(7) & 0.68(2) & 8.75(1)& 0.42(1) \\
\met\ \JK{2}{-1}{1}{-1} E & 0.42(7) & 0.19(2) & fixed  & fixed   && 1.12(7) & 0.54(2) & fixed  & fixed   \\
\met\ \JK{2}{0}{1}{0} E   &$\la0.21$&$\la0.06$&        &         &&$\la0.21$&$\la0.06$&        &         \\
\met\ \JK{3}{0}{2}{0} A$^+$&$\la0.21$&$\la0.06$&       &         && 1.6(2)  & 0.60(3) & 8.79(1)& 0.35(1) \\
\met\ \JK{3}{-1}{2}{-1} E &$\la0.21$&$\la0.06$&        &         && 1.6(2)  & 0.58(3) & fixed  & fixed   \\
CS \J{3}{2}               & 2.5(3)  & 1.37(6) & 8.58(1)& 0.52(2) && 2.4(3)  & 1.65(8) & 8.57(2)& 0.66(3) \\
CS \J{5}{4}               &         &         &        &         && 0.5(2)  & 0.17(3) & 8.75(3)& 0.35(5) \\
SO \JK{3}{2}{2}{1}        & 1.60(7) & 0.80(2) & 8.72(1)& 0.45(1) && 3.18(8) & 1.63(2) & 8.76(1)& 0.48(1) \\
SO \JK{6}{5}{5}{4}        &$\la1.2$ &$\la0.26$&        &         &&$\la1.2$ &$\la0.26$&        &         \\
SO \JK{7}{6}{6}{5}        &$\la2.5$ &$\la0.68$&        &         &&$\la2.5$ &$\la0.55$&        &         \\
\cthd\ \JK{2}{1,2}{1}{0,1}&        &          &        &         && 0.41(6) & 0.17(2) & 8.78(2)& 0.39(4) \\
\cthc\ \JK{5}{0}{4}{0}    &        &          &        &         &&$\la0.19$&$\la0.05$&        &         \\
\htcs\ \JK{4}{0,4}{3}{0,3}&        &          &        &         &&$\la0.36$&$\la0.09$&        &         \\
NS \J{7/2}{5/2} 161.297   &        &          &        &         &&$\la0.45$&$\la0.11$&        &         \\
     \noalign{\smallskip}
     \hline
     \end{tabular}
     \]
     \begin{list}{}{}
\item[$^a$] The $^{13}$CO 1-0 is non-Gaussian. Line  parameters derived 
from the zero (integrated intensity),  first (line velocity) and 
second (line width) order moments of the emission in the 
8.30-9.20 v$_{\rm LSR}$ range (the velocity range of C$^{18}$O emission). 
\item[$^b$] Weak line: fixed line width with the value of the DCO$^+$
2-1 line.
     \end{list}
    \end{table*}
%

%
     \begin{table*}
     \caption[]{Molecular line results for HH 34 C \& D}
     \label{tb:obshh34cd}
     \[
     \begin{tabular}{lcccc@{\hspace{0.1cm}}ccccc}
     \noalign{\smallskip}
     \hline
     \noalign{\smallskip}   
\multicolumn{1}{c}{} &
\multicolumn{5}{c}{HH 34 C} &
\multicolumn{4}{c}{HH 34 D} 
\\
\cline{2-5}\cline{7-10}
\multicolumn{1}{l}{Molecular} &
\multicolumn{1}{c}{$T_{\rm MB}$} &
\multicolumn{1}{c}{$\int T_{\rm MB} dv$} & 
\multicolumn{1}{c}{$v_{\rm LSR}$} &
\multicolumn{2}{c}{$\Delta v_{\rm LSR}$} &
\multicolumn{1}{c}{$T_{\rm MB}$} &
\multicolumn{1}{c}{$\int T_{\rm MB} dv$} & 
\multicolumn{1}{c}{$v_{\rm LSR}$} &
\multicolumn{1}{c}{$\Delta v_{\rm LSR}$} 
\\
\multicolumn{1}{l}{Transition} &
\multicolumn{1}{c}{(K)} &
\multicolumn{1}{c}{$\!\!\!\!\!$(K\kms)} & 
\multicolumn{1}{c}{$\!\!\!\!\!$(km~s$^{-1}$)} &
\multicolumn{2}{c}{$\!\!\!\!\!$(km~s$^{-1}$)} &
\multicolumn{1}{c}{(K)} &
\multicolumn{1}{c}{$\!\!\!\!\!$(K\kms)} & 
\multicolumn{1}{c}{$\!\!\!\!\!$(km~s$^{-1}$)} &
\multicolumn{1}{c}{$\!\!\!\!\!$(km~s$^{-1}$)} 
\\
     \noalign{\smallskip}     
     \hline
     \noalign{\smallskip}     
\tco\ \J{2}{1}$^a$        &11.2(3)  &10.30(7) & 8.53(9)& 0.7(1)  &&11.3(3)  & 9.70(7) & 8.55(9)& 0.7(1)  \\
\hco\ \J{1}{0}            & 3.79(9) & 2.95(2) & 8.51(1)& 0.73(1) && 2.78(9) & 2.22(3) & 8.57(1)& 0.75(1) \\
\hco\ \J{3}{2}            & 2.9(5)  & 1.6(2)  & 8.53(3)& 0.51(9) && 1.5(4)  & 0.93(16)& 8.64(5)& 0.60(11) \\
CS \J{3}{2}               & 2.3(3)  & 1.25(5) & 8.57(1)& 0.52(3) && 2.0(3)  & 1.14(6) & 8.59(1)& 0.55(4) \\
     \noalign{\smallskip}
     \hline
     \end{tabular}
     \]
     \begin{list}{}{}
\item[$^a$] The $^{13}$CO 1-0 is non-Gaussian. Line parameters derived from 
the zero (integrated intensity),  first (line  velocity) and second (line width) order 
moments of the emission in the 8.0-9.0 v$_{\rm LSR}$ range (the velocity range 
of HCO$^+$ emission). 
     \end{list}
    \end{table*}
%


%
     \begin{table*}
     \caption[]{Molecular line results for HH 1}
     \label{tb:obshh1}
     \[
     \begin{tabular}{lcccc}
     \noalign{\smallskip}
     \hline
     \noalign{\smallskip}   
\multicolumn{1}{l}{Molecular} &
\multicolumn{1}{c}{$T_{\rm MB}$} &
\multicolumn{1}{c}{$\int T_{\rm MB} dv$} & 
\multicolumn{1}{c}{$v_{\rm LSR}$} &
\multicolumn{1}{c}{$\Delta v_{\rm LSR}$} 
\\
\multicolumn{1}{l}{Transition} &
\multicolumn{1}{c}{(K)} &
\multicolumn{1}{c}{$\!\!\!\!\!$(K\kms)} & 
\multicolumn{1}{c}{$\!\!\!\!\!$(km~s$^{-1}$)} &
\multicolumn{1}{c}{$\!\!\!\!\!$(km~s$^{-1}$)} 
\\
     \noalign{\smallskip}     
     \hline
     \noalign{\smallskip}     
\tco\ \J{2}{1}$^a$        &16.7(3)   &29.7(2)   & 9.1(1)  & 1.6(2)  \\
\cdo\ \J{2}{1}$^a$        & 5.60(9)  & 6.25(5)  & 9.1(1)  & 1.3(1)  \\
\hco\ \J{1}{0}            & 3.21(10) & 7.10(6)  & 9.21(1) & 2.07(2) \\
\hco\ \J{3}{2}            & 1.6(5)   & 2.50(29) & 9.12(8) & 1.45(20)\\
\htco\ \J{1}{0}           & 0.45(5)  & 0.42(2)  & 9.31(2) & 0.87(5) \\
\htco\ \J{3}{2}           &$\la0.33$ &$\la0.15$ && \\
\dco\ \J{2}{1}            & 0.33(9)  & 0.18(3)  & 9.19(4) & 0.57(13)\\
\dco\ \J{3}{2}            & 0.26(10) & 0.13(3)  & 9.18(7) & 0.45(13)\\
HCN \J{3}{2}              & 0.35(17) & 0.56(11) & 9.14(16)& 1.5(3)  \\
\for\ \JK{2}{0,2}{1}{0,1} & 0.8(2)   & 1.2(2)   & 9.18(9) & 1.5(3)  \\
\met\ \JK{2}{0}{1}{0} A$^+$&0.23(5)  & 0.61(7)  & 9.20(10)& 2.4(2)   \\
\met\ \JK{2}{-1}{1}{-1} E & 0.20(5)  & 0.54(7)  & fixed   & fixed   \\
\met\ \JK{2}{0}{1}{0} E   &$\la0.18$ &$\la0.13$ &&  \\
\met\ \JK{3}{0}{2}{0} A$^+$&$\la0.21$&$\la0.11$ && \\
CS \J{3}{2}               & 1.85(16) & 2.03(11) & 9.09(3) & 1.03(8) \\
SO \JK{3}{2}{2}{1}        & 0.84(3)  &  1.70(2) & 9.05(1) & 1.92(3) \\
SO \JK{6}{5}{5}{4}        &$\la0.6$  &$\la0.28$ && \\
SO \JK{7}{6}{6}{5}        &$\la2.0$  &$\la0.93$ && \\
H$_2$S \JK{1}{1,0}{0}{0,1}& 1.3(3)   & 0.98(15) & 9.01(6) & 0.73(14)\\
     \noalign{\smallskip}
     \hline
     \end{tabular}
     \]
     \begin{list}{}{}
\item[$^a$] The $^{13}$CO 1-0, C$^{18}$O 1- lines are non-Gaussian. Line 
parameters derived from the zero (integrated intensity),  first (line 
velocity) and second (line width) order moments of the emission in the 
8.10-10.30 v$_{\rm LSR}$ range (the velocity range of HCO$^+$ emission). 
     \end{list}
    \end{table*}
%
%

%

     \begin{table*}
     \caption[]{Molecular line results for NGC2264G}
     \label{tb:obsngc}
     \[
     \begin{tabular}{lcccc@{\hspace{0.1cm}}ccccc}
     \noalign{\smallskip}
     \hline
     \noalign{\smallskip}   
\multicolumn{1}{c}{} &
\multicolumn{5}{c}{JMG1} &
\multicolumn{4}{c}{JMG2} 
\\
\cline{2-5}\cline{7-10}
\multicolumn{1}{l}{Molecular} &
\multicolumn{1}{c}{$T_{\rm MB}$} &
\multicolumn{1}{c}{$\int T_{\rm MB} dv$} & 
\multicolumn{1}{c}{$v_{\rm LSR}$} &
\multicolumn{2}{c}{$\Delta v_{\rm LSR}$} &
\multicolumn{1}{c}{$T_{\rm MB}$} &
\multicolumn{1}{c}{$\int T_{\rm MB} dv$} & 
\multicolumn{1}{c}{$v_{\rm LSR}$} &
\multicolumn{1}{c}{$\Delta v_{\rm LSR}$} 
\\
\multicolumn{1}{l}{Transition} &
\multicolumn{1}{c}{(K)} &
\multicolumn{1}{c}{$\!\!\!\!\!$(K\kms)} & 
\multicolumn{1}{c}{$\!\!\!\!\!$(km~s$^{-1}$)} &
\multicolumn{2}{c}{$\!\!\!\!\!$(km~s$^{-1}$)} &
\multicolumn{1}{c}{(K)} &
\multicolumn{1}{c}{$\!\!\!\!\!$(K\kms)} & 
\multicolumn{1}{c}{$\!\!\!\!\!$(km~s$^{-1}$)} &
\multicolumn{1}{c}{$\!\!\!\!\!$(km~s$^{-1}$)} 
\\
     \noalign{\smallskip}     
     \hline
     \noalign{\smallskip}     
\tco\ \J{2}{1}$^0$        & 1.0(2)  & 1.04(12)& fixed  & fixed   && 2.5(3)  & 2.46(7) & 4.86(8)& 0.76(9)\\
                          & 1.7(2)  & 8.5(3)  & 3.42(10)&4.7(2)  && 2.5(3)  & 2.46(7) & 4.86(8)& 0.76(9)\\
                          & 1.4(2)  & 1.39(12)& 8.63(4)& 0.97(9) && 2.5(3)  & 2.46(7) & 4.86(8)& 0.76(9)\\
\cdo\ \J{2}{1}            & 0.61(18)& 0.65(9) & 4.83(7)& 1.01(18)&& 1.04(19)& 0.86(8) & 4.83(4)& 0.77(9)\\
\hco\ \J{1}{0}$^a$        & 0.7(1)  & 2.30(6) & 1.6(4) & 3.0(5)  && 0.4(1)  & 0.33(3) & 4.7(2) & 0.9(2) \\
\hco\ \J{3}{2}$^b$        & 0.7(2)  & 0.6(2)  & 4.9(1) & 0.9(3)  && 2.8(4)  & 2.77(14)& 4.6(1) & 0.8(1 )\\
                          & 0.8(2)  & 2.2(4)  & 2.1(2) & 2.5(6)  && 2.8(4)  & 2.77(14)& 4.6(1) & 0.8(1 )\\
\htco\ \J{1}{0}$^d$       & 0.06(3) & 0.06(2) & 5.1(3) & 0.9(5)  && 0.24(5) & 0.20(2) & 4.64(4)& 0.78(7)\\
\htco\ \J{3}{2}           &$\la0.5$ &$\la0.3$ &        &         &&$\la0.75$&$\la0.29$&& \\
\dco\ \J{2}{1}            &$\la0.15$&$\la0.07$&        &         && 0.15(5) & 0.19(3) & 4.44(9)& 1.1(2) \\
\dco\ \J{3}{2}            &$\la0.18$&$\la0.08$&        &         &&$\la0.27$&$\la0.08$&& \\
HCN \J{3}{2}              & 0.69(14)& 3.87(18)& 2.10(12)& 5.3(3) && 0.84(12)& 2.45(14)& 4.05(7)& 2.7(2) \\
\met\ \JK{2}{0}{1}{0} A$^+$&0.48(2) & 2.78(5) & 2.29(6) & 5.40(6) && 0.56(3) & 1.11(2) & 4.43(2)& 1.83(3)\\
\met\ \JK{2}{-1}{1}{-1} E & 0.35(2) & 2.03(5) & fixed   & fixed   && 0.39(3) & 0.75(2) & fixed  & fixed  \\
\met\ \JK{2}{0}{1}{0} E   & 0.09(2) & 0.53(5) & fixed   & fixed   && 0.07(3) & 0.13(2) & fixed  & fixed  \\
\met\ \JK{2}{1}{1}{1} E   & 0.03(2) & 0.19(5) & fixed   & fixed   &&$\la0.10$&$\la0.03$&& \\
\met\ \JK{3}{0}{2}{0} A$^+$&0.65(10)& 3.64(12)& 1.68(7) & 5.27(9) && 0.73(16)& 1.21(7) & 4.48(5)& 1.56(7)\\
\met\ \JK{3}{-1}{2}{-1} E & 0.60(10)& 3.38(12)& fixed   & fixed   && 0.51(16)& 0.84(8) & fixed  & fixed  \\
\met\ \JK{3}{0}{2}{0}   E &$\la0.28$&$\la0.30$&         &         &&         &         &        &        \\
CS \J{3}{2}               & 0.91(14)& 1.10(14)& 4.88(5) & 1.15(11)&& 1.6(2)  & 2.20(12)&4.58(3) & 1.28(2)\\
                          & 0.78(14)& 4.5(3)  & 1.88(15)& 5.4(3)  &&         &         &        &        \\
SO \JK{3}{2}{2}{1}        & 0.35(5) & 2.21(8) & 1.99(10)& 5.7(2)  && 0.67(7) & 0.88(4) & 4.56(3)& 1.22(7)\\     
SO \JK{6}{5}{5}{4}$^c$    & 0.15(7) & 0.80(10)& 1.6(3)  & 4.7(5)  && 0.36(8) & 0.57(6) & 4.22(7)& 1.4(2) \\     
SO \JK{7}{6}{6}{5}$^c$    & 0.2(1)  & 0.72(18)& 1.6(3)  & 3.5(5)  && 0.3(1)  & 0.17(5) & 4.3(1) & 0.5(2) \\   
H$_2$S \JK{1}{1,0}{0}{0,1}& 0.2(1)  & 0.9(2) & 2.8(4)& 3.9(9) && 0.3(1)  & 0.5(2)  & 4.3(2) & 1.5(4) \\
     \noalign{\smallskip}
     \hline
     \end{tabular}
     \]
     \begin{list}{}{}
\item[$^0$] The $^{13}$CO 1-0 is fitted with three Gaussian components. One is
forced to  have the same velocity peak and line width as the \cdo\ emission. 
\item[$^a$] At the ambient velocity ($\sim 4.8$~\kms) the emission is missing.
Computed the moments for the broader component using the CS velocity range
($-0.6$--4.8~\kms)
\item[$^b$] Used two components fits as in CS \J{3}{2}.
\item[$^d$] Very marginal detection.
\item[$^c$] Marginal detection, so we used the -2.2 to 5.2 \kms\ velocity range
to  derive the integrated line intensities, velocity and line width. 
    \end{list}
    \end{table*}
%

\end{document}